%% file: ANA-HDBS-2018-41-PAPER.tex
\documentclass[cernpreprint, texlive=2016, USenglish, texmf]{atlasdoc}
 
\usepackage{atlaspackage}
\usepackage{atlasbiblatex}
\usepackage[hepparticle]{atlasphysics}
\usepackage{atlasmisc}
 
\graphicspath{{logos/}{figures/}}
 
\usepackage{ANA-HDBS-2018-41-PAPER-defs}
 
\AtlasTitle{Search for resonant pair production of Higgs bosons in the \(b\bar{b}b\bar{b}\) final state using \(pp\) collisions at \(\sqrt{s}=13\;\mathrm{TeV}\) with the ATLAS detector}
 
\AtlasAbstract{
A search for resonant Higgs boson pair production in the \(b\bar{b}b\bar{b}\) final state is presented.
The analysis uses 126--\SI{139}{\ifb} of \(pp\) collision data at \(\sqrt{s} = \SI{13}{\TeV}\) collected with the ATLAS detector at the Large Hadron Collider.
The analysis is divided into two channels, targeting Higgs boson decays which are reconstructed as pairs of small-radius jets or as individual large-radius jets.
Spin-0 and spin-2 benchmark signal models are considered, both of which correspond to resonant \(HH\) production via gluon--gluon fusion.
The data are consistent with Standard Model predictions.
Upper limits are set on the production cross-section times branching ratio to Higgs boson pairs of a new resonance in the mass range from \SI{251}{\GeV} to \SI{5}{\TeV}.
}
 
\author{The ATLAS Collaboration}
 
\AtlasRefCode{HDBS-2018-41}

\AtlasJournalRef{Phys.\ Rev.\ D 105 (2022) 092002}
 
\AtlasDOI{10.1103/PhysRevD.105.092002}
 
\PreprintIdNumber{CERN-EP-2021-229}
\hypersetup{pdftitle={ATLAS document},pdfauthor={The ATLAS Collaboration}}
 
\begin{document}
 
\maketitle
 
\tableofcontents
 
% The next lines are included from the .//sections/intro.tex input file
\section{Introduction}
 
The discovery of the \SI{125}{\GeV} Higgs boson (\(H\))~\cite{HIGG-2012-27,CMS-HIG-12-028,HIGG-2018-04,CMS-HIG-18-016} at the Large Hadron Collider (LHC) enables and motivates the search for resonances decaying into Higgs boson pairs (\(HH\)).
Such searches have been carried out by the ATLAS and CMS collaborations in \(pp\) collisions at center-of-mass energies (\(\sqrt{s}\)) of both \SI{8}{\TeV} and \SI{13}{\TeV}.
All of these have found results that are consistent with the Standard Model (SM) prediction that no such resonances exist.
The ATLAS Collaboration has set constraints on resonant \(HH\) production via vector-boson fusion in the \(b\bar{b}b\bar{b}\) final state~\cite{HDBS-2018-18-custom} and via gluon--gluon fusion (ggF) in the boosted \(b\bar{b}\tau^+\tau^-\) final state~\cite{HDBS-2019-22} using the full Run~2 dataset.
It also searched for resonant \(HH\) production in the \(b\bar{b}b\bar{b}\) decay mode via gluon--gluon fusion~\cite{EXOT-2016-31}, in the \(\gamma\gamma b\bar{b}\)~\cite{HIGG-2016-15}, \(b\bar{b}\tau^+\tau^-\)~\cite{HIGG-2016-16}, \(WWb\bar{b}\)~\cite{HIGG-2016-27}, \(WWWW\)~\cite{HIGG-2016-24}, and \(WW\gamma\gamma\)~\cite{HIGG-2016-20} decay modes, and in a combination~\cite{HDBS-2018-58}, using up to \SI{36.1}{\ifb} of the Run~2 dataset at \(\sqrt{s}=\SI{13}{\TeV}\).
The CMS Collaboration has similarly set constraints in the \(b\bar{b}b\bar{b}\)~\cite{CMS-HIG-17-009}, \(b\bar{b}\gamma\gamma\)~\cite{CMS-HIG-17-008},
\(b\bar{b}\tau^+\tau^-\)~\cite{CMS-HIG-17-002}, \(b\bar{b}\ell\nu\ell\nu\)~\cite{CMS-HIG-17-006}, and \(b\bar{b}ZZ\)~\cite{CMS-HIG-18-013} decay modes,
and a combination of these~\cite{CMS-HIG-17-030}, using \SI{35.9}{\ifb} of \(\sqrt{s}=\SI{13}{\TeV}\) data,
as well as the \(b\bar{b}\tau^+\tau^-\)~\cite{CMS-HIG-20-014} decay mode and final states with \(b\)-quarks and leptons~\cite{CMS-bblep} using the full Run~2 dataset.
 
\begin{figure}[bp]
\centering
\subfloat[]{
\label{fig:feyn-scalar}
\includegraphics[width=0.50\textwidth]{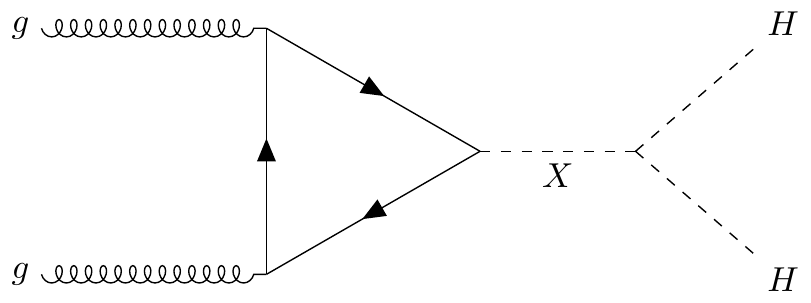}
}
\hspace{0.06\textwidth}
\subfloat[]{
\label{fig:feyn-graviton}
\includegraphics[width=0.36\textwidth]{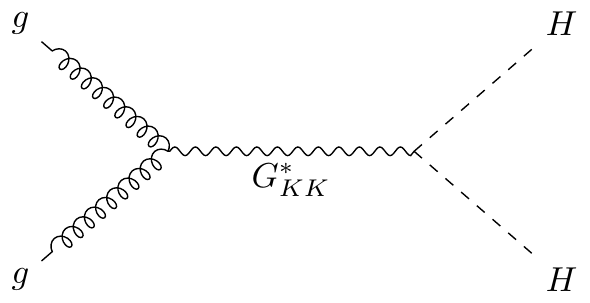}
}
\caption{\label{fig:feyn}
Feynman diagrams for resonant Higgs boson pair production via gluon--gluon fusion in the two benchmark signal models:
\protect \subref{fig:feyn-scalar} a generic spin-0 boson, and \protect \subref{fig:feyn-graviton} a Kaluza--Klein graviton.
}
\end{figure}

This paper presents a search for resonant pair production of Higgs bosons via gluon--gluon fusion in the \(b\bar{b}b\bar{b}\) final state, using the full LHC Run~2 dataset collected by ATLAS.
The results are interpreted in terms of two representative benchmark models: a generic spin-0 boson, \(X\) (as, for example, predicted by two-Higgs-doublet models~\cite{Branco:2011iw} such as the Minimal Supersymmetric Standard Model~\cite{Djouadi:2013vqa,Djouadi:2013uqa}),
and a spin-2 Kaluza--Klein graviton, \GKK, in the context of the bulk Randall--Sundrum (RS) model~\cite{Randall:1999ee,Agashe:2007zd,Fitzpatrick:2007qr,Oliveira:2014kla}.
In both cases, only the gluon--gluon fusion production mode is considered.
No further assumptions are made on the signal models except for the spin hypothesis and generated resonance width.
Example production diagrams for these signal models are shown in Figure~\ref{fig:feyn}.
Throughout this analysis, the nominal \(H \to b\bar{b}\) branching ratio is taken to be 0.582, corresponding to the SM value at a Higgs boson mass of \SI{125}{\GeV}~\cite{deFlorian:2016spz}.
 
The analysis is divided into two complementary channels:
\textit{resolved}, in which each of the four \(b\)-quarks from the \(H\) decays leads to an individually reconstructed jet, and
\textit{boosted}, which targets the topology where each \(H\) is produced with large transverse momentum (\pt) and its decay products are reconstructed as a single large-radius jet.
The resolved and boosted channels target low and high resonance masses, respectively.
The resolved channel covers resonance masses from \SI{251}{\GeV} to \SI{1.5}{\TeV}, and
the boosted channel covers resonance masses from \SI{900}{\GeV} to \SI{5}{\TeV}.
The two channels are statistically combined in the mass range where they overlap.
 
In addition to utilizing the full Run~2 dataset and benefiting from progress in the ATLAS \(b\)-jet identification algorithms, this analysis is improved over the previous ATLAS search~\cite{EXOT-2016-31} in several ways.
In the resolved channel, a machine-learning algorithm is used to pair the jets into Higgs boson candidates
and the fully data-driven background model is substantially improved with a neural-network-based reweighting procedure.
In the boosted channel, variable-radius track-jets are used for \(b\)-tagging to recover signal acceptance for the highest resonance masses,
which fell in the previous analysis as the Higgs boson decay products became very collimated, and the search is extended to cover the previously unexplored resonance mass range between 3 and \SI{5}{\TeV}.
In both channels, the analysis uses a new neural-network-based \(b\)-tagging algorithm DL1r, which performs better than the older MV2 algorithm~\cite{FTAG-2018-01,ATL-PHYS-PUB-2017-013}.
% End of text imported from the .//sections/intro.tex input file
\pagebreak
% The next lines are included from the .//sections/atlas.tex input file
\section{ATLAS detector}
 
The ATLAS detector~\cite{PERF-2007-01} at the LHC covers nearly the entire solid angle around the collision point.\footnote{ATLAS uses a right-handed coordinate system with its origin at the nominal interaction point (IP)
in the center of the detector and the \(z\)-axis along the beam pipe.
The \(x\)-axis points from the IP to the center of the LHC ring,
and the \(y\)-axis points upwards.
Cylindrical coordinates \((r,\phi)\) are used in the transverse plane,
\(\phi\) being the azimuthal angle around the \(z\)-axis.
The pseudorapidity is defined in terms of the polar angle \(\theta\) as \(\eta = -\ln \tan(\theta/2)\).
Angular distance is measured in units of \(\Delta R \equiv \sqrt{(\Delta\eta)^{2} + (\Delta\phi)^{2}}\).}
It consists of an inner tracking detector surrounded by a thin superconducting solenoid, electromagnetic and hadron calorimeters,
and a muon spectrometer incorporating three large superconducting air-core toroidal magnets.
 
The inner-detector (ID) system is immersed in a \SI{2}{\tesla} axial magnetic field and provides charged-particle tracking in the range \(|\eta| < 2.5\).
The high-granularity silicon pixel detector covers the vertex region and typically provides four space-point measurements per track, the first hit normally being in the insertable B-layer installed before Run~2~\cite{ATLAS-TDR-19,PIX-2018-001}.
The next layer outward is the silicon microstrip tracker, which usually provides eight measurements per track.
These silicon detectors are complemented by the transition radiation tracker, which enables radially extended track reconstruction up to \(|\eta| = 2.0\).
 
The calorimeter system covers the pseudorapidity range \(|\eta| < 4.9\).
Within the region \(|\eta|< 3.2\), electromagnetic calorimetry is provided by barrel and endcap high-granularity lead/liquid-argon (LAr) calorimeters, with an additional thin LAr presampler covering \(|\eta| < 1.8\) to correct for energy loss in material upstream of the calorimeters.
Hadron calorimetry is provided by the steel/scintillator-tile calorimeter, segmented into three barrel structures within \(|\eta| < 1.7\), and two copper/LAr hadron endcap calorimeters.
The solid angle coverage is completed with forward copper/LAr and tungsten/LAr calorimeter modules optimized for electromagnetic and hadronic energy measurements respectively.
 
The muon spectrometer (MS) comprises separate trigger and high-precision tracking chambers measuring the deflection of muons in a magnetic field generated by the superconducting air-core toroidal magnets.
The field integral of the toroids ranges between \num{2.0} and \SI{6.0}{\tesla\(\cdot\)\metre} across most of the detector.
A set of precision chambers covers the region \(|\eta| < 2.7\) with three layers of monitored drift tubes, complemented by cathode-strip chambers in the forward region, where the background is highest.
The muon trigger system covers the range \(|\eta| < 2.4\) with resistive-plate chambers in the barrel, and thin-gap chambers in the endcap regions.
 
Interesting events are selected by the first-level trigger system implemented in custom hardware, followed by selections made by algorithms implemented in software in the high-level trigger~\cite{TRIG-2016-01}.
The first-level trigger accepts events from the \SI{40}{\MHz} bunch crossings at a rate below \SI{100}{\kHz},
which the high-level trigger further reduces in order to record events to disk at about \SI{1}{\kHz}.
 
An extensive software suite~\cite{ATL-SOFT-PUB-2021-001} is used in the reconstruction and analysis of real and simulated data, in detector operations, and in the trigger and data acquisition systems of the experiment.
% End of text imported from the .//sections/atlas.tex input file
% The next lines are included from the .//sections/samples.tex input file
\section{Data and simulated samples}
 
\subsection{Data sample}
\label{sec:data}
This analysis is performed using LHC \(pp\) collision data at \(\sqrt{s} = \SI{13}{\TeV}\).
Only data collected during stable beam conditions with all relevant detector systems functional are used~\cite{DAPR-2018-01}.
The resolved (boosted) channel uses \SI{126}{\ifb} (\SI{139}{\ifb}) of data collected in 2016--2018 (2015--2018).
The triggers for both channels are based on jets reconstructed using the \antikt algorithm~\cite{Cacciari:2008gp,Fastjet}.
For the resolved (boosted) channel, the jets are clustered with a radius parameter of \(R=0.4\) (\(R=1.0\)).
 
The resolved channel uses a combination of 12 triggers with various requirements on the transverse energy (\ET) and \(b\)-tagging status of the jets~\cite{TRIG-2018-08}.
These include requirements on any one of four sets of objects:
\begin{itemize}
\item two \(b\)-jets plus two additional jets (2\(b\)+2j),
\item two \(b\)-jets plus one additional jet (2\(b\)+1j),
\item a single high-\ET \(b\)-jet, and
\item two \(b\)-jets plus a high \HT, defined as the scalar sum of all jets' \ET (2\(b\)+\HT).
\end{itemize}
The minimum \ET requirements on the jets are \SI{35}{\GeV} for the \(b\)-jets from the 2\(b\)+2j triggers,
\SI{55}{\GeV} for the \(b\)-jets from the 2\(b\)+\HT triggers,
100 to \SI{150}{\GeV} (depending on the year) for the additional jet used for the 2\(b\)+1j triggers,
and finally 225 or \SI{300}{\GeV} for the single high-\ET \(b\)-jet trigger.
The minimum \HT requirement is \SI{300}{\GeV} for the triggers which include it.
In the trigger, \HT is computed using all jets with \(\ET \geq \SI{30}{\GeV}\).
The efficiency of individual triggers varies from a few percent to up to 80\% depending on the kinematic and \(b\)-tagging requirements of the trigger, and the signal hypothesis.
The choice of triggers is optimized to maximize the signal efficiency over the full range of hypothesized resonance mass values.
The set of triggers used depends on the data-taking year, and the triggers from each year have different effects on kinematic distributions.
This results in a different signal-to-background ratio in each year's data.
Therefore, the datasets from each year are treated independently until they are combined in the statistical analysis.
During 2016 data taking, a fraction of the data (\SI{8.3}{\ifb}) was affected by an inefficiency in the online vertex reconstruction,
which reduced the efficiency of the algorithms used to identify \(b\)-jets; those events were not retained for further analysis.
This results in an integrated luminosity of \SI{24.6}{\ifb} for the 2016 dataset in the resolved channel.
The integrated luminosities of the 2017 and 2018 datasets are \SI{43.7}{\ifb} and \SI{57.7}{\ifb}, respectively.
 
In the boosted channel, events were selected from the 2015 dataset using a trigger that required a single jet, \(J\), with \(\ET > \SI{360}{\GeV}\).
In 2016, 2017 and 2018 a similar trigger was used but requiring \(\ET > \SI{420}{\GeV}\).
The 2017 and 2018 triggers have additional requirements on the mass of the jet of \(m_J > \SI{40}{\GeV}\) and \(m_J > \SI{35}{\GeV}\), respectively.
The efficiency of these triggers is 98\% for data passing the jet requirements described in Section~\ref{sec:boosted},
so the triggers do not significantly impact any relevant kinematic distributions and the datasets corresponding to each year are combined into one dataset.
 
\subsection{Simulated samples}
 
Monte Carlo (MC) simulation is used for the modeling of signal events and, in the case of the boosted channel, for the modeling of the \ttbar background.
The ATLAS detector response is simulated with \GEANT~\cite{Agostinelli:2002hh} for background samples, spin-0 signal samples with a resonance mass of \SI{1}{\TeV} or higher, and all spin-2 signal samples.
\textsc{AtlFastII}~\cite{SOFT-2010-01}, which utilizes a fast calorimeter simulation, is used for spin-0 signals with resonance masses below \SI{1}{\TeV}.
 
The signal processes for both benchmark models were simulated at leading order (LO) in \(\alphas\), using \MGNLO[2.6.1]~\cite{Alwall:2014hca} for the spin-0 samples and \MGNLO[2.2.2] for the spin-2 samples.
For both cases, the \NNPDF[2.3lo]~\cite{Ball:2012cx} parton distribution function (PDF) set was used.
Both resonances were produced via gluon--gluon fusion and were forced to decay into a pair of SM Higgs bosons, as indicated in Figure~\ref{fig:feyn}.
Signal samples were generated for resonance masses in a range from \SI{251}{\GeV} to \SI{5}{\TeV}, with increased spacing between the higher mass-points.
The spacing becomes larger than the detector resolution at high masses.
For the spin-0 model, interpolation methods~\cite{Morphing} are used to estimate the mass distribution between the points above \SI{2}{\TeV} for which samples were simulated.
The spin-2 resonance is wide compared to the mass spacing, so no interpolation is required.
 
For the spin-0 case, a two-Higgs-doublet model was used in the event generation, taking the heavy CP-even neutral scalar as the resonance of interest.
Its width was set to be much smaller than the detector resolution, and the other non-SM particles in this model do not enter the generation
(At LO in QCD, there are no Feynman diagrams for this process containing them).
Hadronization and parton showering were modeled with \HERWIG[7.1.3]~\cite{Bellm:2017jjp}, using the \MMHT[lo]~\cite{Harland-Lang:2014zoa-custom}
PDF set for the parton shower and \EVTGEN[1.6.0]~\cite{Lange:2001uf} to model heavy-flavor decays.
The \HERWIG[7.1] default set of tuned underlying-event parameters was used.
No theoretical cross-sections are required for this process; the simulated events are used purely to model the kinematics.
 
For the spin-2 case, a Kaluza--Klein graviton with \(k/\overline{M}_{\mathrm{Pl}}\) = 1 is taken as the benchmark,
where \(k\) is the curvature of the warped extra dimension and \(\overline{M}_{\mathrm{Pl}} = 2.4 \times 10^{18}~\GeV\)~is the effective four-dimensional Planck scale.
The generated width, based on the model prediction, ranges from 3\% to 20\% of the resonance mass, which is not negligible compared to the detector resolution.
The hadronization and showering were modeled using \PYTHIA[8.186]~\cite{Sjostrand:2014zea} with \EVTGEN[1.2.0] for heavy-flavor decays.
The A14 set of tuned underlying-event parameters~\cite{ATL-PHYS-PUB-2014-021} and the \NNPDF[2.3lo] PDF set were used.
The cross-sections for this process are taken from Ref.~\cite{Oliveira:2014kla}.
They are used solely for setting limits on the graviton mass.
 
Top quark pair production (\ttbar) was simulated at next-to-leading order (NLO) in \(\alphas\) using \POWHEGBOX[v2]~\cite{Nason:2004rx,Frixione:2007vw,Alioli:2010xd,Campbell:2014kua}.
Parton showering, hadronization, and the underlying event were modeled using \PYTHIA[8.230] with \EVTGEN[1.6.0] for heavy-flavor decays.
The matrix element calculation uses \NNPDF[3.0nlo]~\cite{Ball:2014uwa} as the PDF set,
while the parton shower and underlying-event modeling uses \NNPDF[2.3lo]~\cite{Ball:2012cx} and the A14 set of tuned parameters.
The damping parameter \hdamp, which effectively regulates radiation at high \pt, was set to 1.5 times the top quark mass.
The \ttbar simulation is normalized using the value of the inclusive cross-section calculated with \TOPpp[2.0]~\cite{Czakon:2011xx,ATL-PHYS-PUB-2018-009}.
This accounts for next-to-next-to-leading-order (NNLO) corrections in \(\alphas\), including next-to-next-to-leading logarithmic (NNLL) resummation of soft gluon terms.
 
Multijet background processes were modeled using \PYTHIA[8.235].
This simulates pure quantum chromodynamics (QCD) 2-to-2 interactions at LO in \(\alphas\).
Events were showered using the parton shower native to \PYTHIA, which includes radiation and splitting that can result in additional jets, along with \EVTGEN[1.6.0] for heavy-flavor decays.
The A14 set of tuned underlying-event parameters and the \NNPDF[2.3lo] PDF set were used.
 
Other background processes, such as SM \(H\), \(HH\), and electroweak diboson production, have been estimated to give negligible contributions to the selected event yields and are therefore not included.
 
The effect of multiple interactions in the same and neighboring bunch
crossings (pileup) was modeled by overlaying each simulated hard-scattering event with
inelastic proton--proton ($pp$) events generated with \PYTHIA[8.186]
using the \NNPDF[2.3lo] PDF and the
A3 set of tuned parameters~\cite{ATL-PHYS-PUB-2016-017}.
% End of text imported from the .//sections/samples.tex input file
% The next lines are included from the .//sections/objects.tex input file
\section{Object reconstruction}
 
Primary vertices from proton--proton interactions are reconstructed using at least two charged-particle tracks with \(\pt > \SI{500}{\MeV}\) measured with the ID~\cite{PERF-2015-01}.
The vertex which has the largest sum of squared track momenta (\(\sum \pt^2)\) is selected as the hard-scatter primary vertex.
 
Hadronic jets are reconstructed using the \antikt algorithm~\cite{Cacciari:2008gp,Fastjet}.
Depending on the use case, different input objects and radius parameters \(R\) are used.
 
For the resolved channel, \textit{small-R jets} are clustered using \(R=0.4\) with particle-flow objects as inputs~\cite{PERF-2015-09}.
Particle-flow objects are charged-particle tracks matched to the hard-scatter vertex
and calorimeter energy clusters following an energy subtraction algorithm that removes the calorimeter deposits associated with good-quality tracks from any vertex.
The tracking information is used to improve the clusters' energy resolutions.
The momenta of these jets are calibrated in a multi-step procedure~\cite{JETM-2018-05}.
Jets with \(\pt < \SI{60}{\GeV}\) and \(|\eta| < 2.4\) must also satisfy a requirement based on the output of the multivariate ``jet vertex tagger'' (JVT)
algorithm, which is used to identify and reject jets in which much of the energy originates from pileup interactions~\cite{PERF-2014-03}.
The ``Tight'' working point, corresponding to an average signal efficiency of 96\%, is used and jets failing this requirement are discarded.
All small-\(R\) jets are required to have \(\pt > \SI{40}{\GeV}\) and \(|\eta| < 2.5\).
Any jets failing these requirements are discarded and not used further, except where stated explicitly.
 
Additional small-\(R\) jets are reconstructed from topological clusters of energy deposits in the calorimeter~\cite{PERF-2014-07} instead of particle-flow objects.
These jets are used exclusively for the purpose of applying quality criteria to identify events which are consistent with noise in the calorimeter or noncollision background~\cite{ATLAS-CONF-2015-029}.
They are calibrated in the same way as the small-\(R\) jets reconstructed from particle-flow objects.
If an event contains at least one jet which has \(\pt > \SI{20}{\GeV}\), passes the JVT, and fails to meet these quality criteria, the event is rejected.
 
For the boosted channel, \textit{large-R jets} are clustered using \(R=1.0\) with topological clusters of energy deposits in calorimeter cells as the input.
The clusters are locally calibrated~\cite{PERF-2011-03} before being combined into jets.
After these large-\(R\) jets are created, a trimming procedure~\cite{Krohn:2009th} is applied to mitigate the effects of pileup:
the constituents are reclustered into ``subjets'' using the \(k_t\) algorithm~\cite{Ellis:1993tq} with \(R=0.2\),
and any of these subjets with less than 5\% of the large-\(R\) jet's \pt are removed.
The large-\(R\) jets are calibrated following a procedure similar to that for the small-\(R\) jets;
however, there is no area-based pileup subtraction step or global sequential calibration~\cite{PERF-2012-02}.
Additionally, the mass of each large-\(R\) jet is calibrated using both calorimeter and track information~\cite{ATLAS-CONF-2016-035}.
 
The boosted channel also makes use of \textit{track-jets} to identify individual \(b\)-hadron decays within the large-\(R\) jet.
ID tracks are clustered using the \antikt algorithm with a variable radius.
The effective radius \(R\) is inversely proportional to the \pt of the constituent(s) in question: \(R = \rho/\pt\).
Here, a value of \(\rho = \SI{30}{\GeV}\) is used.
Minimum and maximum values of this effective radius are set at \(R_\mathrm{min} = 0.02\) and \(R_\mathrm{max} = 0.4\).
Track-jets do not have a dedicated calibration; their momenta are taken to be the vector sum of the momenta of their constituent tracks.
After being reconstructed, these track-jets are exclusively matched to large-\(R\) jets using the ghost association method~\cite{Cacciari:2007fd}.
 
A \(b\)-tagging algorithm~\cite{FTAG-2018-01,ATL-PHYS-PUB-2017-013} is applied to both the small-\(R\) jets and the track-jets to identify those which are likely to have originated from a \(b\)-quark.
The DL1r algorithm is used, at a working point chosen to have 77\% efficiency on average for jets associated with true \(b\)-hadrons in simulated \ttbar events.
This is a multivariate algorithm which uses a selection of inputs including information about the impact parameters of ID tracks, the presence of displaced secondary vertices, and the reconstructed flight paths of \(b\)- and \(c\)-hadrons inside the jet~\cite{ATL-PHYS-PUB-2017-003}.
At the chosen working point, the light-jet (charm-jet) rejection measured in \ttbar MC simulation is about a factor of 130 (4.9) on average for small-\(R\) jets.
The training and calibration of this algorithm is performed separately for each jet type~\cite{ATLAS-CONF-2018-001,ATLAS-CONF-2018-006}.
Correction factors are applied to the simulated samples to compensate for differences between the \(b\)-tagging efficiencies in data and simulation.
 
Muons are reconstructed by matching ID tracks with either MS tracks or aligned individual hits in the MS and performing a combined track fit.
They are required to have \(\pt > \SI{4}{\GeV}\) and \(|\eta| < 2.5\), and to satisfy ``Medium'' identification criteria based on track-quality variables~\cite{MUON-2018-03}.
Muons are used only to apply corrections to jet momenta.
 
A momentum correction is applied to \(b\)-tagged small-\(R\) jets to account for energy lost to soft out-of-cone radiation and to muons and neutrinos in semileptonic \(b\)-hadron decays.
This correction follows the procedure used in Ref.~\cite{HIGG-2016-29} and consists of two parts.
For the first, if any muon is within \(\Delta R = 0.4\) of a \(b\)-tagged small-\(R\) jet, the four-momentum of the muon is added to that of the jet.
Any energy deposited in the calorimeter by the muon is then subtracted from the jet to prevent double counting; this is computed according to the description in Ref.~\cite{PERF-2015-10}.
In the second step a global scale factor is applied to each \(b\)-tagged small-\(R\) jet based on its \pt and whether or not it has a muon within \(\Delta R = 0.4\) of the jet axis.
These scale factors are derived from simulation.
 
To account for energy lost in semileptonic \(b\)-hadron decays, a similar muon-in-jet correction is applied to large-\(R\) jets.
If a muon is matched within a distance of  \(\Delta R = \mathrm{min}\left(0.4,0.04+10\;\GeV / p_\mathrm{T}^\mathrm{muon}\right)\) to one of the two leading track-jets associated with the large-\(R\) jet, and if the track-jet is \(b\)-tagged, the muon is considered part of the large-\(R\) jet.
Again, any energy deposited in the calorimeter by the muon is subtracted from the jet to prevent double counting.
The muon four-momentum is then added to the calorimeter-based component of the large-\(R\) jet four-momentum, and the jet mass is recalculated~\cite{PERF-2017-04}.
 
% End of text imported from the .//sections/objects.tex input file
% The next lines are included from the .//sections/resolved.tex input file
\section{Resolved channel}
 
\subsection{Event selection}
 
To be considered for analysis, events must pass the trigger requirements specified in Section~\ref{sec:data}.
To simplify the modeling of trigger efficiencies, events are sorted into exactly one of four classes based on offline kinematic quantities,
each of which requires one specific type of trigger to be passed.
In decreasing order of priority, these are as follows:
\begin{enumerate}
\item If the leading jet has \(\pt > \SI{325}{\GeV}\) and is \(b\)-tagged, the trigger requiring one high-\ET \(b\)-jet is used.
\item If the leading jet has \(\pt > \SI{170}{\GeV}\) and is not \(b\)-tagged, the trigger requiring two \(b\)-jets and one additional jet is used.
\item If the \HT in the event (computed using all jets with \(\pt > \SI{25}{\GeV}\) and \(|\eta| < 2.5\)) is greater than \SI{900}{\GeV},
the trigger requiring a high \HT is used.
\item For all remaining events, the trigger requiring two \(b\)-jets and two additional jets is used.
\end{enumerate}
The definitions of these classes are the result of a dedicated sensitivity optimization intended to minimize the expected limits on the signal cross-section.
Following the trigger selection, events are required to have at least four small-\(R\) jets.
Events are then divided into two categories, ``\(2b\)'' (where exactly two jets are \(b\)-tagged) and ``\(4b\)'' (where at least four jets are \(b\)-tagged).
Exactly four jets are selected to construct the two \(H\) candidates.
For \(4b\) events, the four \(b\)-tagged jets with the highest \pt are selected.
For \(2b\) events, the two \(b\)-tagged jets and the two untagged jets with the highest \pt are selected.
The \(2b\) events are needed to construct the background model for the \(4b\) category.
This selection of untagged jets can introduce a kinematic bias with respect to the \(4b\) category, but this is accounted for by the reweighting function described in Section~\ref{sec:res-bkg}.
 
After the four jets are chosen, there are three possible combinations for pairing them into \(H\) candidates.
For a given pairing, the four-momentum of the \(H\) candidate is defined as the sum of the four-momenta of the jets used to construct it.
The pairing is chosen using a boosted decision tree (BDT).
This is trained using LightGBM~\cite{LightGBM} to classify each of the three possible pairings in a signal event as either correct or incorrect.
The correct pairing is defined by using the generator's decay record to match jets to the parton-level \(b\)-quarks which result immediately from the \(H\) decay.
The classifier assigns each of the three candidate pairings a score, and the pairing with the highest score is chosen.
The input variables to the BDT are the separations in pseudorapidity , azimuthal angle, and their quadratic sum (\(\Delta\eta\), \(\Delta\phi\), and \(\Delta R\), respectively) between the two jets in each pair.
Although the same information is contained in \(\Delta\eta\) and \(\Delta\phi\) alone,
additional use of a precalculated \(\Delta R\) leads to improved performance.
The BDT is also parameterized in the invariant mass of the four identified \(b\)-jets (\(m(4b)\)), which is included as an additional input feature.
However, \(m(4b)\) cannot itself be used to discriminate between correct and incorrect pairings as it is independent of the pairing.
Its inclusion as a parameter serves the purpose of ensuring optimal performance for all resonance masses.
The BDT is trained on a sample consisting of one quarter of the simulated spin-0 signal events, across the full range of resonance masses considered in the resolved channel.
A further selection on training events is applied, requiring them to contain at least four jets with \(\pt > \SI{35}{\GeV}\),
each of which is within \(\Delta R = 0.4\) of a true \(b\)-quark, defined at parton level and originating from a Higgs boson decay.
This \pt requirement is loosened relative to the nominal selection in order to increase the number of selected events with low \(m(4b)\).
The events entering this training are not used anywhere else in the analysis.
This technique results in less background in the signal region (defined below) compared to the strategy used in the previous ATLAS search~\cite{EXOT-2016-31},
which was based on minimizing the difference between the two di-jet invariant masses.
The BDT algorithm finds the correct pairing in at least 65\% of signal events.
This efficiency is lowest for low resonance masses, but reaches almost 100\% for resonance masses of \SI{600}{\GeV} and higher.
This is a significant improvement with respect to the efficiency of the previous method, especially at low resonance masses.
 
After the \(H\) candidates are formed, they are ordered by the scalar sum of the \pt of their constituent jets: \(H_1\) and \(H_2\) denote the leading and subleading \(H\) candidates, respectively.
A pseudorapidity separation between the two \(H\) candidates of \(|\Delta\eta_{HH}|<1.5\) is required in order to reduce the multijet background.
Additionally, a ``top veto'' is applied, to reduce the background from hadronic top quark decays.
This is defined by combining every possible pair of jets with \(\pt > \SI{40}{\GeV}\) and \(|\eta| < 2.5\), including those that were not selected for the \(H\) candidates, to form ``\(W\) candidates''.
``Top quark candidates'' are built by pairing \(W\) candidates with each remaining jet that was selected for \(H\) candidates.
Events are rejected if any top quark candidate satisfies \(X_{Wt} < 1.5\), where the discriminant \(X_{Wt}\) is defined as
 
\begin{equation}\nonumber
X_{Wt} = \sqrt{\left(\frac{m(W) - m_\textrm{SM}(W)}{0.1 \times m(W)}\right)^2
+ \left(\frac{m(t) - m_\textrm{SM}(t)}{0.1 \times m(t)}\right)^2}.
\end{equation}
 
Here, \(m(W)\) and \(m(t)\) denote the masses of the candidates under consideration, while \(m_\textrm{SM}(W)\) and \(m_\textrm{SM}(t)\) denote the measured masses of these particles
(\SI{80.4}{\GeV} and \SI{172.5}{\GeV}, respectively~\cite{Zyla:2020zbs}).
The denominators in the expression for \(X_{Wt}\) represent the approximate mass resolution of the detector.
 
Finally, events are sorted into three kinematic regions based on the invariant masses of the \(H\) candidates.
The first of these is the \textit{signal region} (SR), defined by requiring \(X_{HH} < 1.6\), where
 
\begin{equation}\nonumber
X_{HH} = \sqrt{\left(\frac{m(H_1) - \SI{120}{\GeV}}{0.1 \times m(H_1)}\right)^2
+ \left(\frac{m(H_2) - \SI{110}{\GeV}}{0.1 \times m(H_2)}\right)^2}.
\end{equation}
 
\noindent
The shape of the SR in the \(m(H_1)\)--\(m(H_2)\) plane is chosen to optimize the signal significance.
The mass values of \SI{120}{\GeV} and \SI{110}{\GeV} correspond to the position of the peaks of the simulated signal \(m(H_1)\) and \(m(H_2)\) distributions.
The deviations from the measured Higgs boson mass of \SI{125}{\GeV}~\cite{Zyla:2020zbs} are due to detector effects, as well as energy lost to neutrinos from the \(b\)-hadron decays and to out-of-cone radiation.
Jets which lose energy give rise to lower mass in their \(H\) candidate.
Since these jets are more likely to compose \(H_2\) by definition, this results in a slightly lower mass for \(H_2\) than \(H_1\) on average.
The \textit{validation region} (VR) contains the events not in the SR which satisfy the condition
 
\begin{equation}\nonumber
\Rhh^{\text{VR}} \equiv \sqrt{\left(m(H_1) - 1.03 \times \SI{120}{\GeV}\right)^2 + \left(m(H_2) - 1.03 \times \SI{110}{\GeV}\right)^2} < \SI{30}{\GeV}.
\end{equation}
 
\noindent
Finally, the \textit{control region} (CR) contains the events not in the SR or VR which satisfy the condition
\begin{equation}\nonumber
\Rhh^{\text{CR}} \equiv \sqrt{\left(m(H_1) - 1.05 \times \SI{120}{\GeV}\right)^2 + \left(m(H_2) - 1.05 \times \SI{110}{\GeV}\right)^2} < \SI{45}{\GeV}.
\end{equation}
 
\begin{figure}[tbp]
\centering
\includegraphics[width=0.65\textwidth]{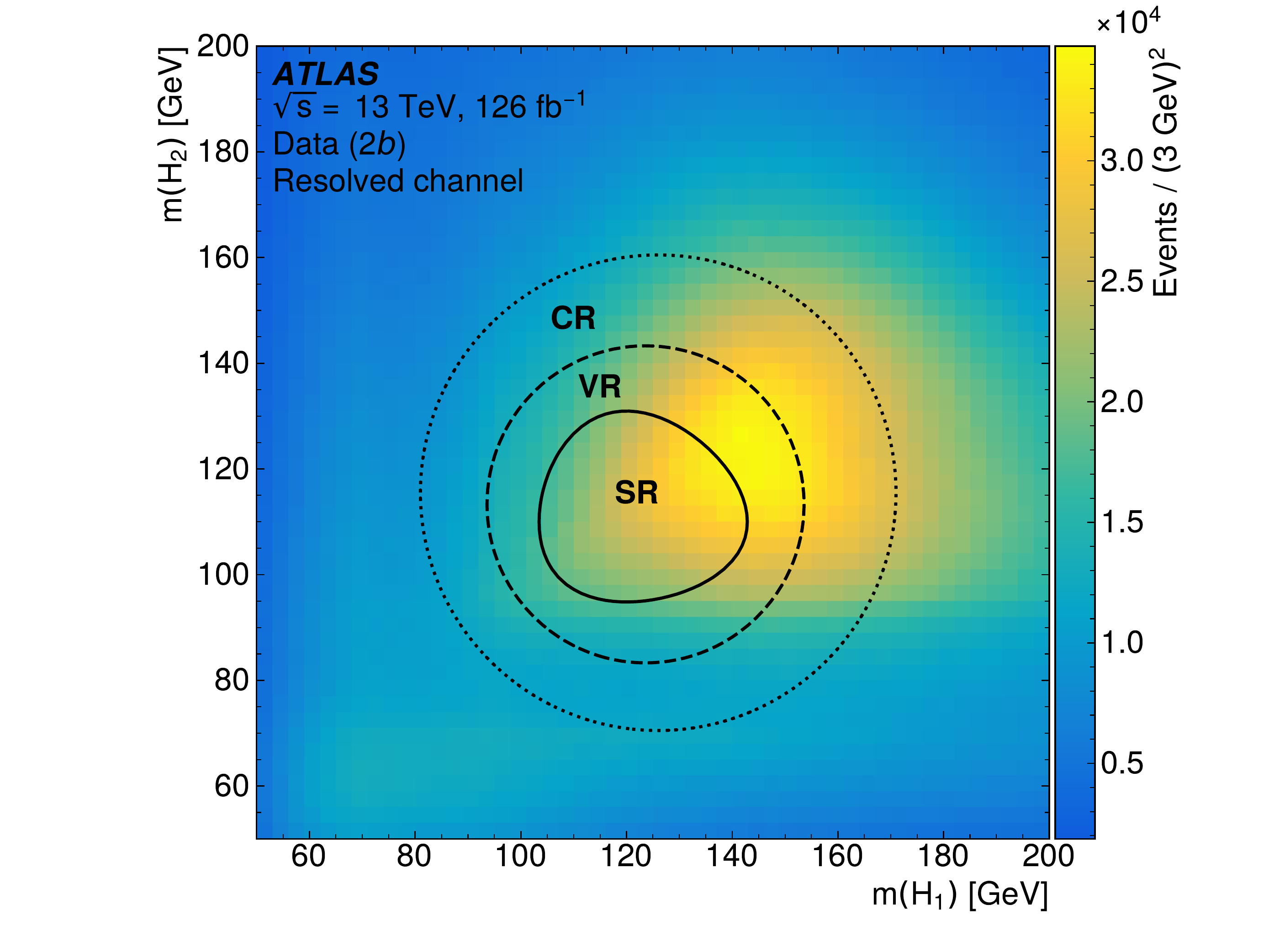}
\caption{
\label{fig:region-def-resolved}
Kinematic region definitions superimposed on the resolved \(2b\) data for the full 2016--2018 dataset. \(H_1\) and \(H_2\) are the reconstructed Higgs boson candidates, sorted by their \(p_\mathrm{T}\).
}
\end{figure}

\noindent
The centers of the VR and CR are shifted relative to that of the SR to ensure that the mean \(H\) candidate masses are equal in the three regions.
The shapes of these regions in the \(m(H_1)\)--\(m(H_2)\) plane are shown with the \(2b\) data in Figure~\ref{fig:region-def-resolved}.
 
\begin{figure}[tbp]
\centering
\subfloat[]{
\label{fig:resolved-acc-scal}
\includegraphics[width=0.48\textwidth]{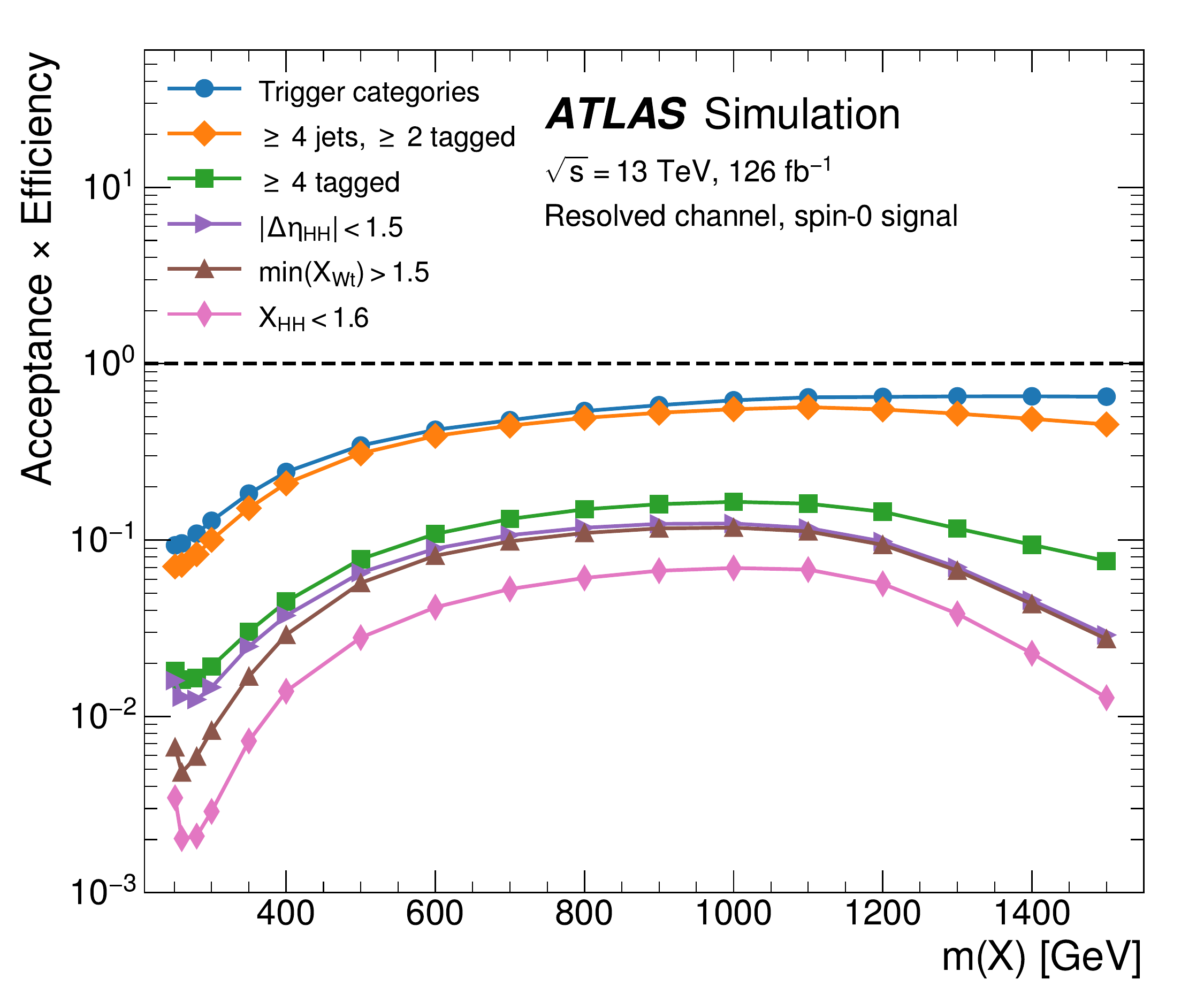}
}
\subfloat[]{
\label{fig:resolved-acc-grav}
\includegraphics[width=0.48\textwidth]{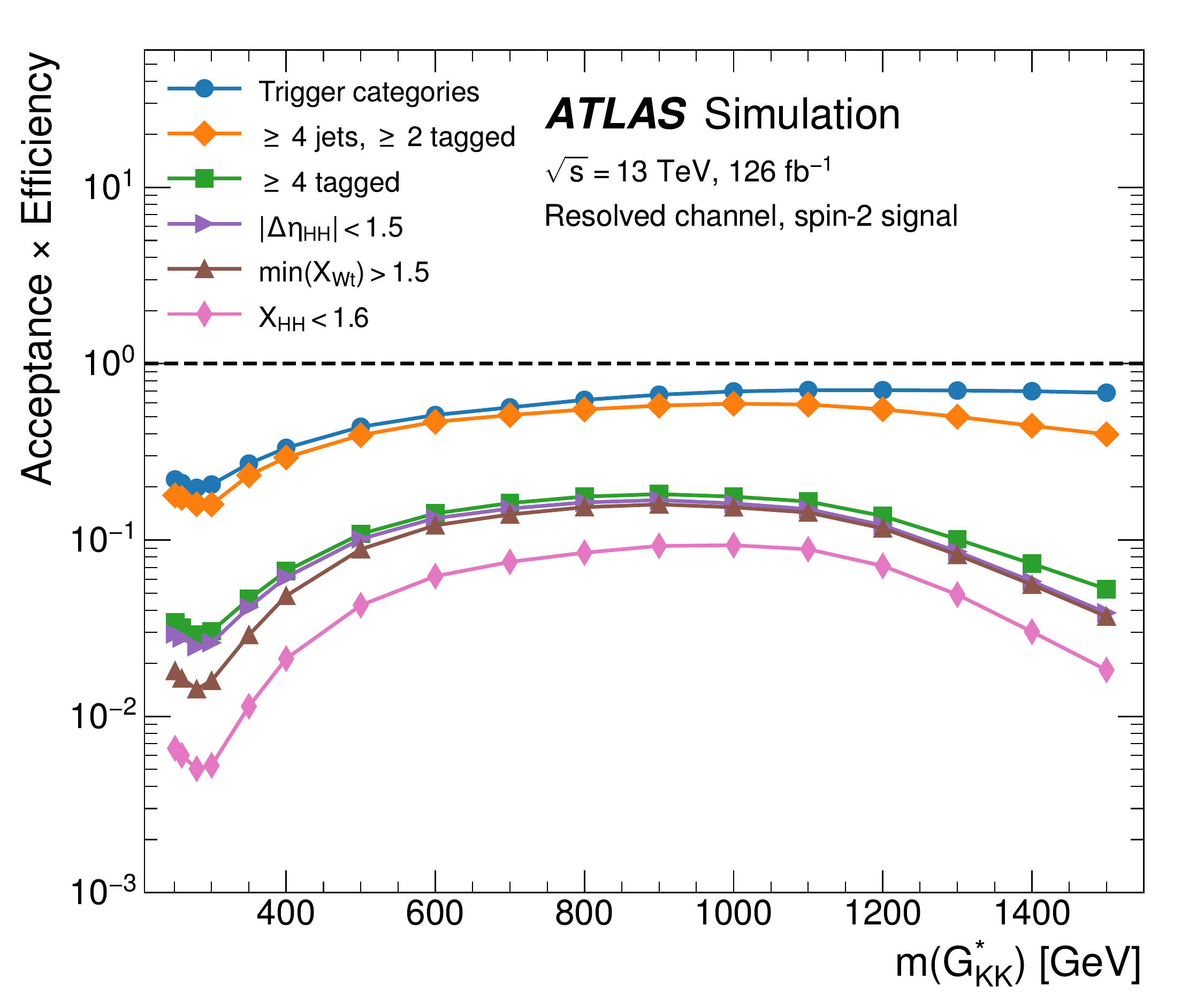}
}
\caption{ \label{fig:resolved-acceptance}
Cumulative acceptance times efficiency as a function of resonance mass for each event selection step in the resolved channel for
\protect \subref{fig:resolved-acc-scal} the spin-0 and
\protect \subref{fig:resolved-acc-grav} the spin-2 signal models.
The local maximum at \SI{251}{\GeV} is a consequence of the near-threshold kinematics.
}
\end{figure}

After the full selection, the final discriminating variable ``corrected \mhh'' is constructed.
This is obtained by rescaling the four-momenta of the \(H\) candidates such that \(m(H_1) = m(H_2) = \SI{125}{\GeV}\).
The corrected \mhh is then the invariant mass of the sum of the two resulting four-momenta.
This procedure improves the scale and resolution of the reconstructed signal mass distribution by correcting for detector effects and physical processes such as radiative emission outside the jet cones.
This correction improves the signal mass resolution by up to 25\% and shifts the mean of the mass distribution closer to the true value.
It also modifies the background shape, but does not introduce any signal-like features.
The signal efficiency times acceptance for the various event selection steps is shown in Figure~\ref{fig:resolved-acceptance}.
The efficiency at low resonance masses is mainly limited by the trigger.
At high resonance masses the jets start to merge together and the reconstruction and \(b\)-tagging efficiencies decrease.
The efficiency is substantially larger for the spin-2 model than for the spin-0 model because the corrected \mhh distribution of the spin-2 model is much broader.
This has an especially large effect at the lowest resonance masses, where the shape of the corrected \mhh distribution is distorted toward higher values.
This is a result of both the \mhh correction and the natural mass shape of the graviton resonance near the production threshold.
 
\subsection{Background estimation}
\label{sec:res-bkg}
 
After the selection described above, the background is dominated by pure QCD multijet processes (excluding top quark production), with the approximately 5\% remainder almost entirely composed of \ttbar.
This background composition was determined by comparing \ttbar MC simulation to the total background estimate in the SR;
it is purely meant to be indicative and is not used in the statistical treatment.
The background is modeled using a purely data-driven technique; the only MC simulation used is for the signal.
 
The background shape in the \(4b\) SR is estimated from data in the \(2b\) regions and the \(4b\) CR.
The signal contamination in the \(2b\) dataset is evaluated and found to be negligible compared to the background uncertainties.
The signal-to-background ratio in the \(4b\) CR depends on the signal hypothesis and ranges from 10\% to 25\% of that in the SR.
However, the impact on the analysis results was studied and found negligible.
This was determined by injecting various signals into the CR (and VR) data using cross sections corresponding to existing experimental upper limits.
 
The event kinematics in the \(2b\) and \(4b\) regions are not expected to be identical, so a reweighting function which maps the \(2b\) kinematic distributions onto the \(4b\) distributions is derived:
 
\begin{equation} \label{eq:reweight}
w(\vec{x}) = \frac{p_{4b}(\vec{x})}{p_{2b}(\vec{x})},
\end{equation}
 
\noindent where \(p_{2b}(\vec{x})\) and \(p_{4b}(\vec{x})\) are the probability density functions for \(2b\) and \(4b\) data, respectively, over a set of kinematic variables \(\vec{x}\).
This function is derived in the CR and then applied to the \(2b\) SR in order to produce a model of the background in the \(4b\) SR.
It can also be applied more generally to any \(2b\) region to produce a background model for the corresponding \(4b\) region.
 
The computation of \(w(\vec{x})\) is a density ratio estimation problem, for which a variety of approaches exist.
The method employed in this analysis is modified from Refs.~\cite{NNloss2,NNLoss} and makes use of an artificial neural network (NN).
This NN is trained on \(2b\) and \(4b\) CR data to minimize the loss function:
 
\begin{equation}\nonumber
\mathcal{L}(w(\vec{x})) = \int d\vec{x} \left[\sqrt{w(\vec{x})}p_{2b}(\vec{x})
+ \frac{1}{\sqrt{w(\vec{x})}}p_{4b}(\vec{x})\right].
\end{equation}
 
\noindent The function in Eq.~(\ref{eq:reweight}) optimizes this loss by equalizing the contributions from the two terms.
 
The kinematic variables used to make up \(\vec{x}\) are chosen to be sensitive to the differences between the \(2b\) and \(4b\) events.
These are:
\begin{enumerate}
\item \(\log(\pt)\) of the selected jet with the second-highest \pt,
\item \(\log(\pt)\) of the selected jet with the fourth-highest \pt,
\item \(\log(\Delta R)\) between the two selected jets with the smallest \(\Delta R\),
\item \(\log(\Delta R)\) between the other two selected jets,
\item the average \(|\eta|\) of selected jets,
\item \(\log(\pt)\) of the \(HH\) system,
\item \(\Delta R\) between the two \(H\) candidates,
\item \(\Delta \phi\) between the jets making up \(H_1\),
\item \(\Delta \phi\) between the jets making up \(H_2\),
\item \(\log(\min(X_{Wt}))\), and
\item the number of jets in the event with \(\pt > \SI{40}{\GeV}\) and \(|\eta| < 2.5\), including jets that are not selected.
\end{enumerate}
 
\noindent Here, ``selected'' jets refer to the four jets which are used to construct the \(H\) candidates.
The variables to which the reweighting is most sensitive are the jet multiplicity, \(\Delta R\) between the two \(H\) candidates, and \(\log(\pt)\) of the \(HH\) system.
The NN has three densely connected hidden layers of 50 nodes, each with a rectified linear unit activation function~\cite{Nair:2010}, and a single-node linear output.
 
The training of the NN is subject to variation due to initial conditions and the limited size of the training samples.
To account for these effects, the bootstrap resampling technique is used~\cite{Bootstrap}.
This entails constructing a set of training samples by sampling with replacement from the original.
The NN is trained independently on each element of this set, using different initial conditions each time.
This results in an ensemble of background estimates.
Since the original training sample is large, the resulting background estimate in each bin can be approximated as being Gaussian-distributed.
Additionally, this sampling-with-replacement procedure can be approximated by applying a randomly distributed integer weight to each event, drawn from a Poisson distribution with a mean of 1.
Both of these approximations are used in order to reduce the computational complexity of the problem.
These ``bootstrap weights'' are independent of \(w(\vec{x})\), which reweights the nominal \(2b\) kinematic distributions to the nominal \(4b\) kinematic distributions.
To increase the stability of the background estimate, the median value of \(w(\vec{x})\) for each event is calculated across the ensemble and used as the nominal background estimate.
This ensemble of weights is also used to evaluate the uncertainty due to the finite training sample, as detailed in Section~\ref{sec:resolvedSyst}.
 
\begin{figure}[tbp]
\centering
\subfloat[]{
\label{fig:resolved-mhh-cr-norw}
\includegraphics[width=0.48\textwidth]{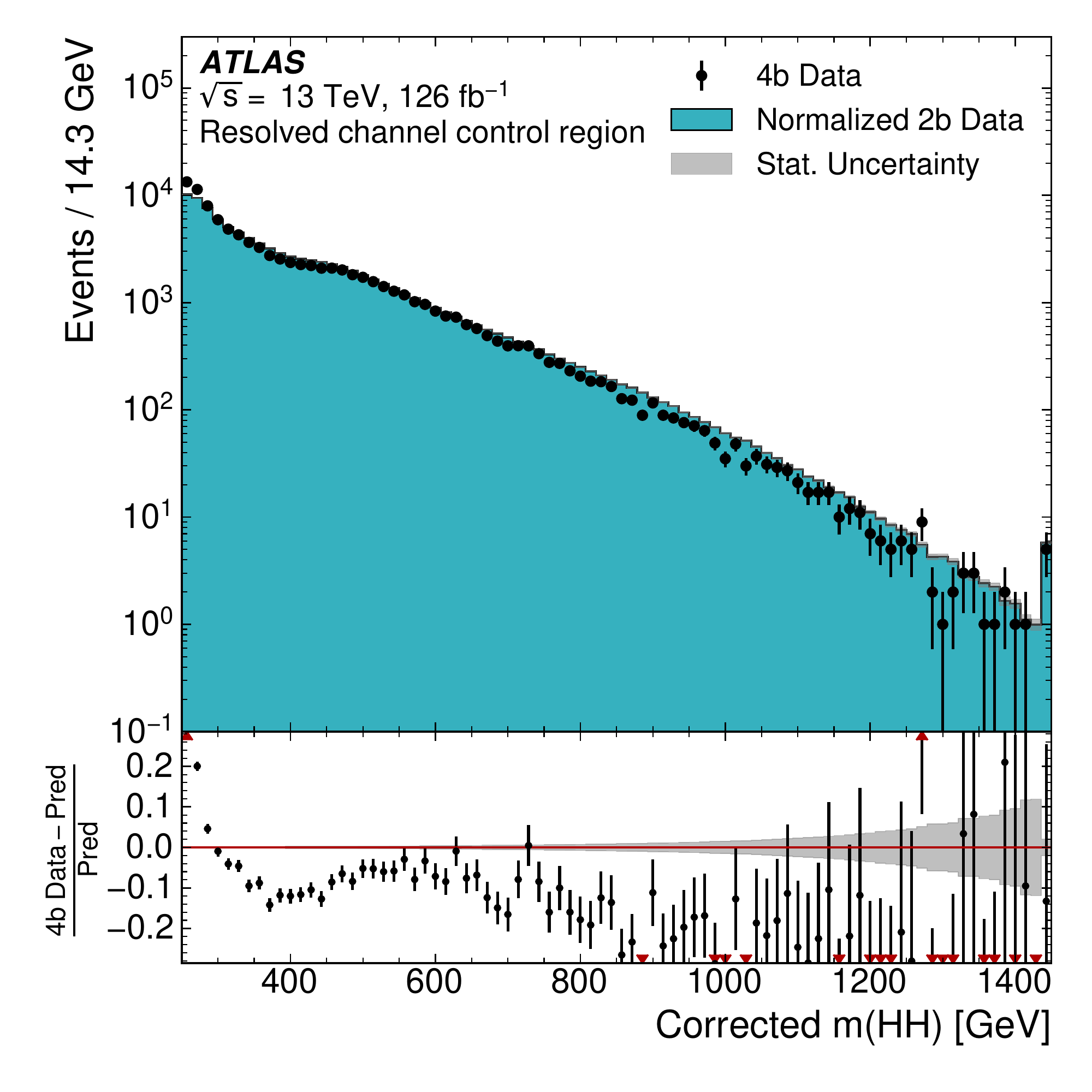}
}
\subfloat[]{
\label{fig:resolved-mhh-cr-rw}
\includegraphics[width=0.48\textwidth]{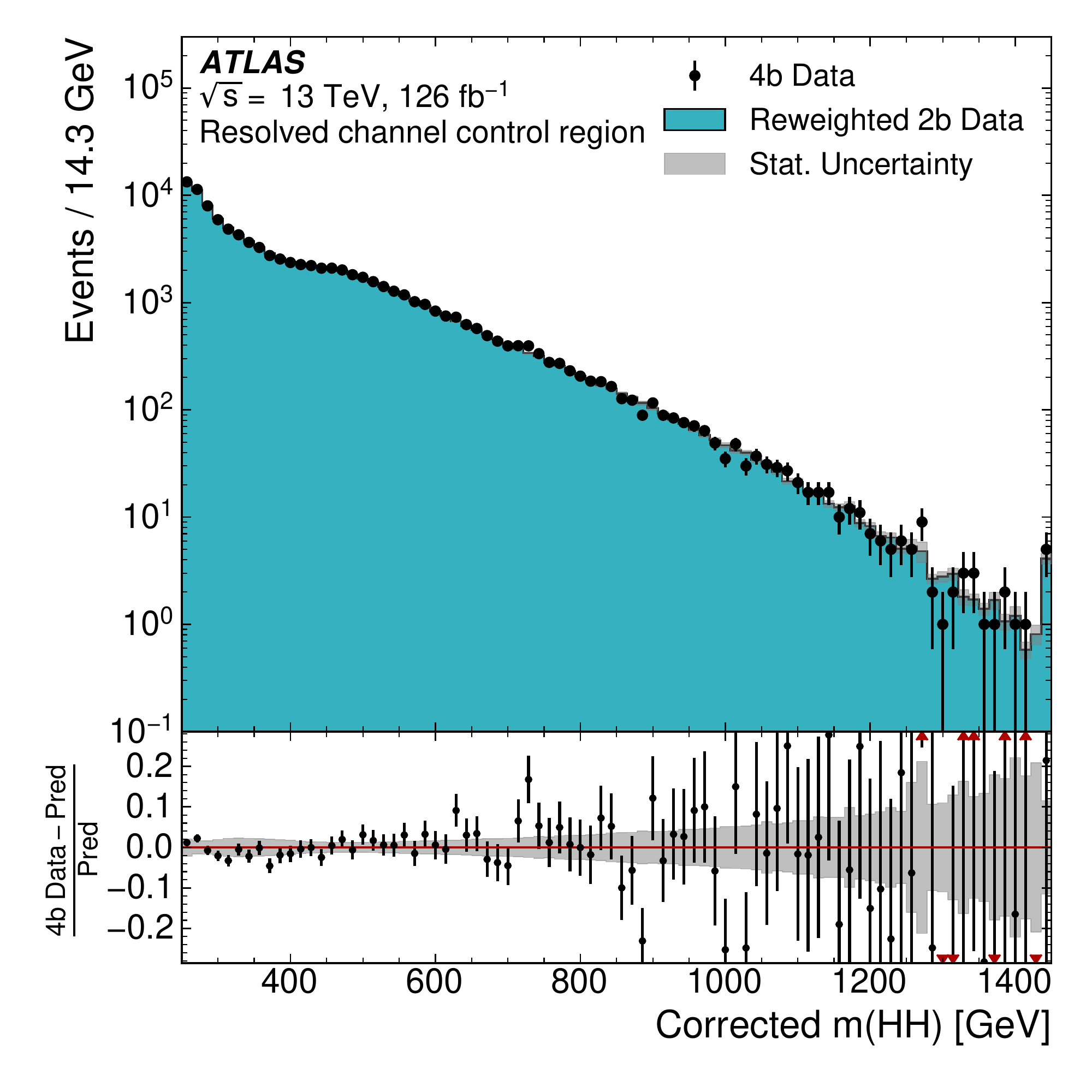}
}
\caption{
\label{fig:resolved-mhh-reweighting}
Corrected \mhh distributions for the \(2b\) control region (teal histogram) and \(4b\) control region (dots) in the resolved channel.
The statistical uncertainty in the \(2b\) control region is represented by the gray band.
The error bars on the \(4b\) points represent the Poisson uncertainties corresponding to their event yields.
The \(2b\) data are shown \protect\subref{fig:resolved-mhh-cr-norw} before and \protect\subref{fig:resolved-mhh-cr-rw} after the kinematic reweighting procedure.
In both cases the \(2b\) distributions are normalized to the \(4b\) event yields for a pure shape comparison.
The final bin of each distribution includes overflow.
The bottom panel shows the difference between the \(4b\) and \(2b\) distributions, relative to the \(2b\) distribution.
}
\end{figure}

\begin{figure}[tbp]
\centering
\includegraphics[width=0.48\textwidth]{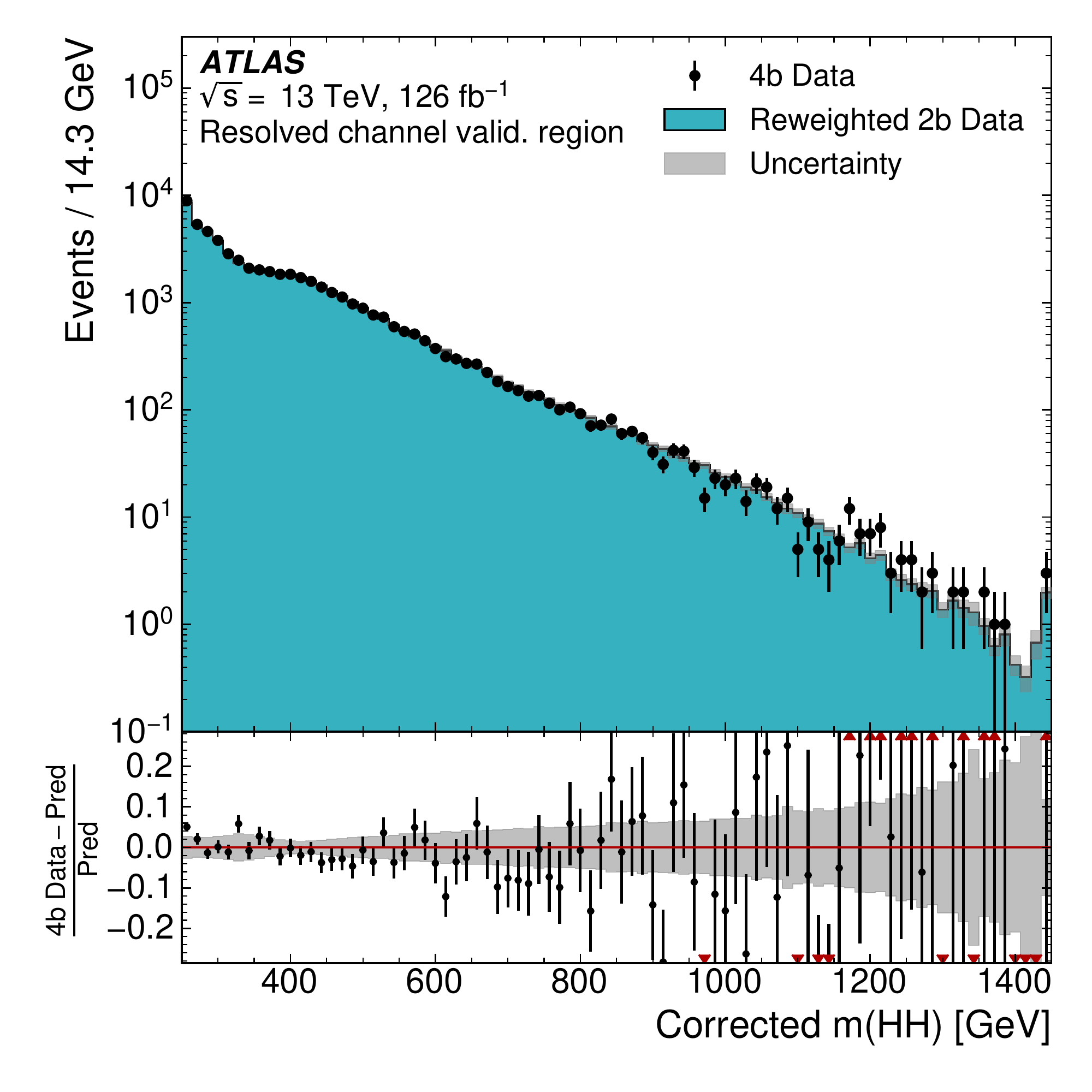}
\caption{
\label{fig:resolved-mhh-vr}
Corrected \mhh distribution in the resolved \(4b\) validation region (dots), compared with the reweighted distribution in \(2b\) validation region (teal histogram).
The error bars on the \(4b\) points represent the Poisson uncertainties corresponding to their event yields.
The final bin includes overflow.
The background uncertainty (gray band) is computed by adding all individual components in quadrature.
The bottom panel shows the difference between the \(4b\) and reweighted \(2b\) distributions, relative to the \(2b\) distribution.
}
\end{figure}

The effect of applying this reweighting to the CR, where it is derived, is shown in Figure~\ref{fig:resolved-mhh-reweighting}.
The output of this procedure is an estimate of the corrected \mhh distribution in the \(4b\) SR,
which is then used as input to the statistical procedure detailed in Section~\ref{sec:results}.
The optimization of the bin width of the corrected \mhh distribution is based on the detector resolution at low masses.
The performance of this reweighting outside of the region where it is derived is checked using the VR.
The scaling of the \(2b\) distribution to the \(4b\) sample size is always derived from the two respective CRs, as part of the reweighting procedure.
The data are found to be compatible with the background model in the VR, as shown in Figure~\ref{fig:resolved-mhh-vr}.
Residual differences between the CR and VR are used to estimate a systematic uncertainty as described in Section~\ref{sec:resolvedSyst}.
 
\subsection{Systematic uncertainties}
\label{sec:resolvedSyst}
 
The most limiting uncertainties in the resolved channel are those arising from the data-driven background estimate which is derived in the CR and applied in the SR.
There are two main sources: uncertainties from the limited sample size in the CR, and physical differences between the CR and SR.
 
The limited sample size in the CR can lead to a random bias in the reweighting function \(w(\vec{x})\).
Ideally, this effect would be evaluated using the event-level covariance matrix between all bootstrap weights; however, this is computationally intractable in practice.
Instead, an approximation is computed using the interquartile range (IQR) of each event's weight distribution as well as the IQR of the normalization factor \(\alpha^{(j)}\)
for each bootstrap training \(j\), which is defined as
 
\begin{equation}\nonumber
\alpha^{(j)} = \frac{n_{4b}}{\sum_{i \in 2b} w_i^{(j)}},
\end{equation}
 
\noindent where \(n_{4b}\) is the number of \(4b\) events, \(w_i^{(j)}\) is the weight for event \(i\) from the bootstrap resampling \(j\),
and the \(2b\) and \(4b\) datasets are restricted to the region in which the reweighting NNs are trained.
Varied distributions are constructed by assigning each event a weight which is varied to the upper boundary of the event-level weight IQR.
These varied distributions are then scaled to have the same normalization as the nominal distribution,
multiplied by the ratio of the upper boundary value of the IQR of \(\alpha^{(j)}\) to its nominal value.
These rescaled varied distributions form an envelope around the nominal one, which specifies the size of this uncertainty in each corrected \mhh bin.
As this uncertainty is statistical in origin, it is uncorrelated across \mhh bins.
 
Uncertainties in the background estimate also arise from kinematic differences between the CR and the SR.
To evaluate these effects, an alternative background model is derived in the VR instead of the CR.
The difference between the corrected \mhh distributions from the nominal and alternative background models is used to estimate the uncertainty in the shape of the \mhh distribution in the SR.
To allow sufficient flexibility in the model, this uncertainty is parameterized in terms of two components: low-\HT and high-\HT,
where \HT now denotes the scalar sum of the \pt of the four jets constituting the \(H\) candidates.
This variable is chosen because it is correlated with \mhh, but does not introduce a discontinuity in the \mhh spectrum when the two components are varied separately.
The boundary between low-\HT and high-\HT events is chosen to be \SI{300}{\GeV}.
Each of these two components is symmetrized around the nominal shape to construct a two-sided uncertainty.
These uncertainties are taken to be uncorrelated across the different years to accommodate differences due to the varying triggers and run conditions.
 
Several detector modeling uncertainties are evaluated and included.
These affect only the signal description, as the background is estimated entirely from data.
Uncertainties in the jet energy scale and resolution are treated according to the prescription in Ref.~\cite{JETM-2018-05}.
Uncertainties in the \(b\)-tagging efficiency are treated according to the prescription in Ref.~\cite{FTAG-2018-01}.
Uncertainties in the trigger efficiencies are evaluated from measurements of per-jet online efficiencies for both jet reconstruction and \(b\)-tagging, which are used to compute event-level uncertainties.
These are then applied to the simulated events as overall weight variations.
The uncertainty in the total integrated luminosity used in this analysis is 1.7\%~\cite{ATLAS-CONF-2019-021},
obtained using the LUCID-2 detector for the primary luminosity measurements~\cite{Avoni:2018iuv}.
 
Several sources of theoretical uncertainty affecting the signal models were considered and are described as follows.
Uncertainties due to modeling of the parton shower and underlying event are evaluated by comparing results between two generators for these parts of the calculation:
the nominal \HERWIG[7.1.3] and the alternative \PYTHIA[8.235].
This is found to have a 5\% effect on the signal acceptance and a negligible impact on the \mhh distribution, independently of the resonance mass.
Uncertainties in the matrix element calculation are evaluated by raising and lowering the factorization and renormalization scales used in the generator by a factor of two, both independently and simultaneously.
This results in an effect smaller than 1\% for all variations and all masses; the impact of such uncertainties is therefore neglected.
PDF uncertainties are evaluated using the PDF4LHC\_NLO\_MC set~\cite{Butterworth:2015oua} by calculating the signal acceptance for each replica and taking the standard deviation.
In all cases, these result in a less than 1\% uncertainty in the signal acceptance, and therefore these are also neglected.
Theoretical uncertainties in the \(H \to b\bar{b}\) branching ratio~\cite{deFlorian:2016spz} are included; they amount to a 2.5\% overall uncertainty on the signal normalization.
 
\subsection{Results}
 
\begin{figure}[tbp]
\centering
\includegraphics[width=0.7\textwidth]{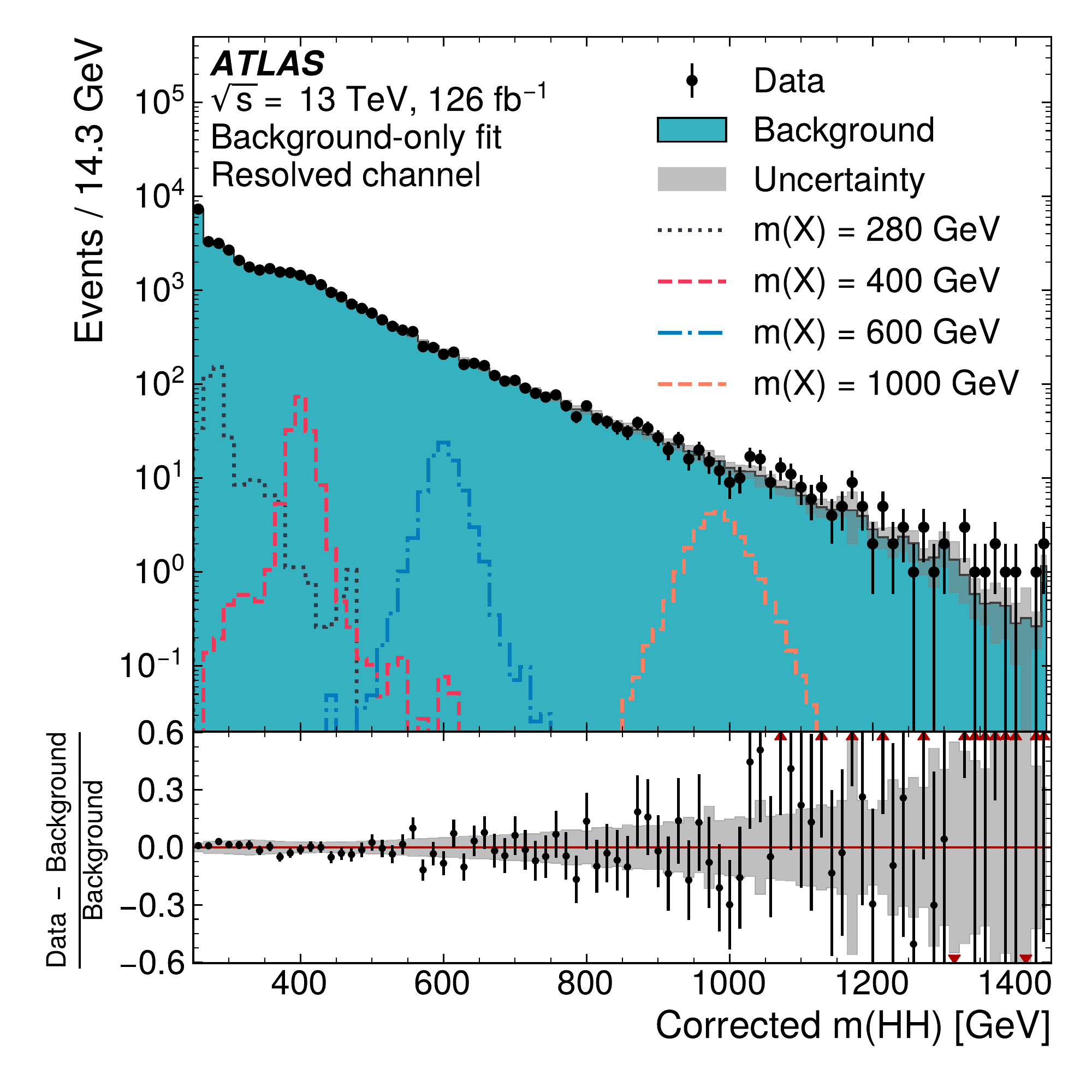}
\caption{
\label{fig:resolved-mhh}
Corrected \mhh distribution in the resolved \(4b\) signal region (dots), after the fit under the background-only hypothesis.
The error bars on the \(4b\) points represent the Poisson uncertainties corresponding to their event yields.
The background model (teal histogram) is shown with its total post-fit uncertainty (gray band).
The final bin includes overflow.
Representative spin-0 signal hypotheses (dashed, dotted, and dashed-dotted lines) are overlaid, normalized to the overall expected limits on their cross-sections.
The bottom panel shows the difference between the \(4b\) distribution and the background model, relative to the background model.
No significant excess of data relative to the SM background is observed.
}
\end{figure}

The corrected \mhh distributions for data and the estimated background after the fit to data described in Section~\ref{sec:results} are shown in Figure~\ref{fig:resolved-mhh}.
The data agree well with the background prediction and no significant excess is observed.
The event yields for data, background, and several signal hypotheses are presented in Tables~\ref{tab:resolved-yields-spin0} and \ref{tab:resolved-yields-spin2}.
These are integrated over windows in the corrected \mhh spectrum containing approximately 90\% of the signal in each case.
These windows are defined such that their first and last bins contain the \(5^\text{th}\) and \(95^\text{th}\) percentile of the distribution, respectively.
This range is larger for the spin-2 signal because the benchmark model predicts a wider resonance.
The statistical interpretation of the data is discussed in Section~\ref{sec:results}.
 
\begin{table}[tbp]
\begin{center}
\caption{Resolved \(4b\) signal region data, estimated background, and signal event yields in corrected \mhh windows containing roughly 90\% of each signal, for representative spin-0 mass hypotheses.
The signal is normalized to the overall expected limit on its cross-section; its uncertainties are evaluated by adding all individual components in quadrature.
The background yields and uncertainties are evaluated after a background-only fit to the data.
}
\label{tab:resolved-yields-spin0}
\sisetup{group-minimum-digits=5}
\begin{tabular}{cc S[table-format=5.0] S[table-format=5.1]@{\(\;\pm\;\)}S[table-format=3.1] S[table-format=5.2]@{\(\;\pm\;\)}S[table-format=2.2]}
\toprule
\(m(X)\) [\GeV] & Corrected \mhh range [\GeV] & \multicolumn{1}{c}{Data}  & \multicolumn{2}{c}{Background model} & \multicolumn{2}{c}{Spin-0 signal model} \\
\midrule
260  &   [250, 321] & 18554 & 18300   & 110   & 503   & 43 \\
500  &   [464, 536] &  2827 &  2866   &  22   & 105.4 & 5.7 \\
800  &   [750, 850] &   358 &   366.2 &   7.3 &  37.7 & 1.7 \\
1200~~ & [1079, 1250] &    68 &    52.6 &   1.7 &  11.71 & 0.62 \\
\bottomrule
\end{tabular}
\end{center}
\end{table}
 
\begin{table}[tbp]
\begin{center}
\caption{Resolved \(4b\) signal region data, estimated background, and signal event yields in corrected \mhh windows containing roughly 90\% of each signal, for representative spin-2 mass hypotheses.
The signal is normalized to the overall expected limit on its cross-section; its uncertainties are evaluated by adding all individual components in quadrature.
The background yields and uncertainties are evaluated after a background-only fit to the data.
}
\label{tab:resolved-yields-spin2}
\sisetup{group-minimum-digits=5}
\begin{tabular}{cc S[table-format=5.0] S[table-format=5.1]@{\(\;\pm\;\)}S[table-format=3.1] S[table-format=5.2]@{\(\;\pm\;\)}S[table-format=2.2]}
\toprule
\(m(\GKK)\) [\GeV] & Corrected \mhh range [\GeV] & \multicolumn{1}{c}{Data}  & \multicolumn{2}{c}{Background model} & \multicolumn{2}{c}{Spin-2 signal model} \\
\midrule
260  &  [250, 393] & 26775 & 26650   & 130   & 368    & 25 \\
500  &  [464, 636] &  4655 &  4719   &  37   & 138.6  &  5.7 \\
800  &  [707, 950] &   795 &   811   &  13   &  52.1  &  1.9 \\
1200~~ & ~~[993, 1279] &   146 &   120.6 &   2.8 &  14.45 &  0.67 \\
\bottomrule
\end{tabular}
\end{center}
\end{table}
 
\FloatBarrier
% End of text imported from the .//sections/resolved.tex input file
% The next lines are included from the .//sections/boosted.tex input file
\section{Boosted channel}
\label{sec:boosted}
 
\subsection{Event selection}
 
After passing the trigger requirement, each event is required to contain at least two large-\(R\) jets with \(\pt > \SI{250}{\GeV}\).
The two highest-\pt jets are selected as the \(H\) candidates.
The leading and subleading \(H\) candidates ordered by \pt are denoted by \(H_1\) and \(H_2\), respectively.
Each \(H\) candidate is required to have \(|\eta| < 2.0\), \(m(H) > \SI{50}{\GeV}\), and at least one associated track-jet.
Additionally, at least one \(H\) candidate must have \(\pt > \SI{450}{\GeV}\), which is driven by the trigger threshold.
In order to further reduce the background, an additional requirement of \(|\Delta\eta_{HH}| < 1.3\) is placed on the \(H\) candidates.
 
Events are then categorized according to the multiplicity and \(b\)-tagging status of the track-jets associated with each of the two \(H\) candidates.
For \(H\) candidates with more than two associated track-jets, only the two with the highest \pt are considered.
 
Since track-jets have a variable radius, it is possible for a high-\pt jet to be contained completely within the catchment area of another low-\pt jet. This can lead to a pathological case in which the low-\pt jet's axis is also contained within the high-\pt jet.
This can result in misassignment of tracks to jets by the \(b\)-tagging algorithm (which is based purely on proximity to the jet axis).
To avoid any degradation in performance resulting from this, events containing such collinear track-jets are vetoed from the boosted channel.
 
Three signal-enriched categories with four or fewer \(b\)-tagged track-jets are defined:
 
\begin{itemize}
\item Events in the ``4\(b\)'' category have two \(b\)-tagged track-jets associated with each \(H\) candidate.
\item Events in the ``3\(b\)'' category have two \(b\)-tagged track-jets associated with one \(H\) candidate and exactly one \(b\)-tagged track-jet associated with the other \(H\) candidate.
\item Events in the ``2\(b\)'' category have exactly one \(b\)-tagged track-jet associated with each \(H\) candidate.
\end{itemize}
These are collectively labeled as the \textit{high-tag} categories.
Including events with less than four \(b\)-tagged track-jets increases the sensitivity of the search especially for high resonance masses, where, due to the large boost, track-jets can become so close that they are often not reconstructed individually.
 
\noindent Additional \textit{low-tag} categories with track-jets that fail the \(b\)-tagging requirement are also defined in order to estimate the background.
 
\begin{figure}[tbp]
\centering
\includegraphics[width=0.9\textwidth]{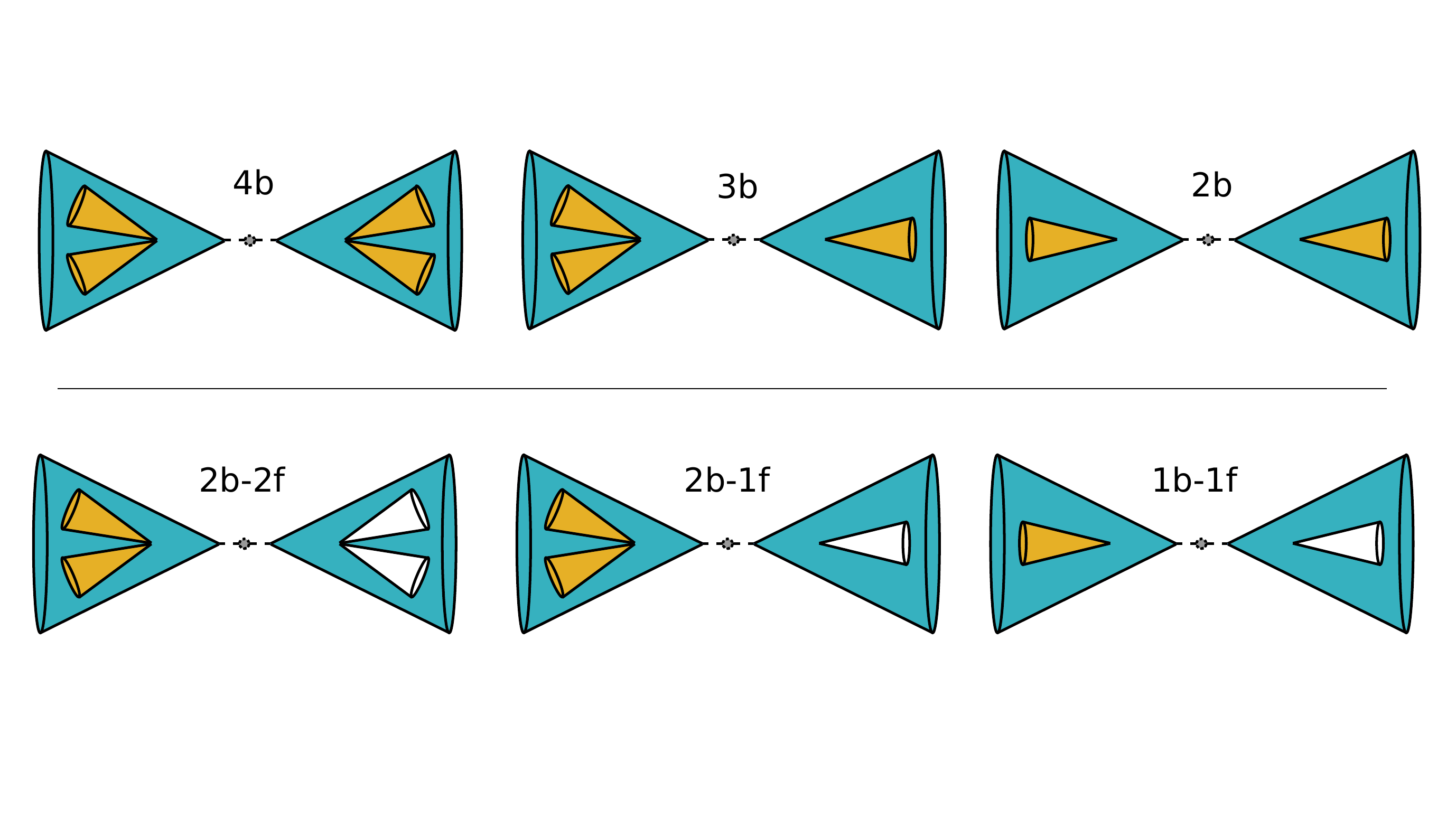}
\caption{
\label{fig:boosted-btag-cats}
Illustration of the three high-tag categories (4\(b\), 3\(b\), and 2\(b\)) with the corresponding low-tag categories
used to estimate the multijet background (2\(b\)-2\(f\), 2\(b\)-1\(f\), and 1\(b\)-1\(f\)).
Teal cones represent large-\(R\) jets, yellow cones represent associated \(b\)-tagged track-jets, and white cones represent associated untagged track-jets.
For \(H\) candidates with more than two associated track-jets, only the two with the highest \pt are considered.
}
\end{figure}

\begin{itemize}
\item Events in the ``2\(b\)-2\(f\)'' category (for modeling 4\(b\)) have one \(H\) candidate with two or more associated \(b\)-tagged track-jets and the other \(H\) candidate with no \(b\)-tagged track-jets but two or more untagged track-jets.
\item Events in the ``2\(b\)-1\(f\)'' category (for modeling 3\(b\)) have one \(H\) candidate with two or more associated \(b\)-tagged track-jets and the other \(H\) candidate with no \(b\)-tagged track-jets but one or more untagged track-jets.
\item Events in the ``1\(b\)-1\(f\)'' category (for modeling 2\(b\)) have one \(H\) candidate with exactly one associated \(b\)-tagged track-jet and the other \(H\) candidate with no \(b\)-tagged track-jets but one or more untagged track-jets.
\end{itemize}
In these low-tag categories, the \(H\) candidate that has no \(b\)-tagged track-jets is also referred to as \textit{untagged}, while the other one is labeled as \textit{tagged}.
The untagged \(H\) candidate in the 2\(b\)-1\(f\) region is allowed to have more than one track-jet because requiring exactly one would result in a very small number of events in this category.
A diagram of events in these high-tag and low-tag categories is shown in Figure~\ref{fig:boosted-btag-cats}.
 
\noindent Events satisfying the 2\(b\)-2\(f\) criteria also necessarily satisfy the 2\(b\)-1\(f\) criteria.
To avoid overlap between the two categories, these events are distributed randomly between them,
with 80\% allocated as 2\(b\)-1\(f\) events and the remaining 20\% allocated as 2\(b\)-2\(f\) events.
This corresponds roughly to the ratio of background events present in the two categories.
 
Similarly to the resolved channel, events are sorted into signal, validation, and control regions based on the invariant masses of the \(H\) candidates.
The SR is defined by requiring \(X_{HH} < 1.6\), where
 
\begin{equation}\nonumber
X_{HH} = \sqrt{\left(\frac{m(H_1) - \SI{124}{\GeV}}{0.1 \times m(H_1)}\right)^2
+ \left(\frac{m(H_2) - \SI{115}{\GeV}}{0.1 \times m(H_2)}\right)^2}.
\end{equation}
 
\noindent
This definition, as well as those of the validation and control regions, slightly differs from that in the resolved channel.
This is due to the different energy scale of the boosted jet reconstruction and the different background distribution.
The VR contains the events not in the SR which satisfy the condition
 
\begin{equation}\nonumber
\Rhh^{\text{VR}} \equiv \sqrt{\left(m(H_1) - \SI{124}{\GeV}\right)^2 + \left(m(H_2) - \SI{115}{\GeV}\right)^2} < \SI{33}{\GeV}.
\end{equation}
 
\noindent
Finally, the CR contains the events not in the SR or VR which satisfy the condition
\begin{equation}\nonumber
\Rhh^{\text{CR}} \equiv \sqrt{\left(m(H_1) - \SI{134}{\GeV}\right)^2 + \left(m(H_2) - \SI{125}{\GeV}\right)^2} < \SI{58}{\GeV}.
\end{equation}
 
\begin{figure}[tbp]
\centering
\includegraphics[width=0.65\textwidth]{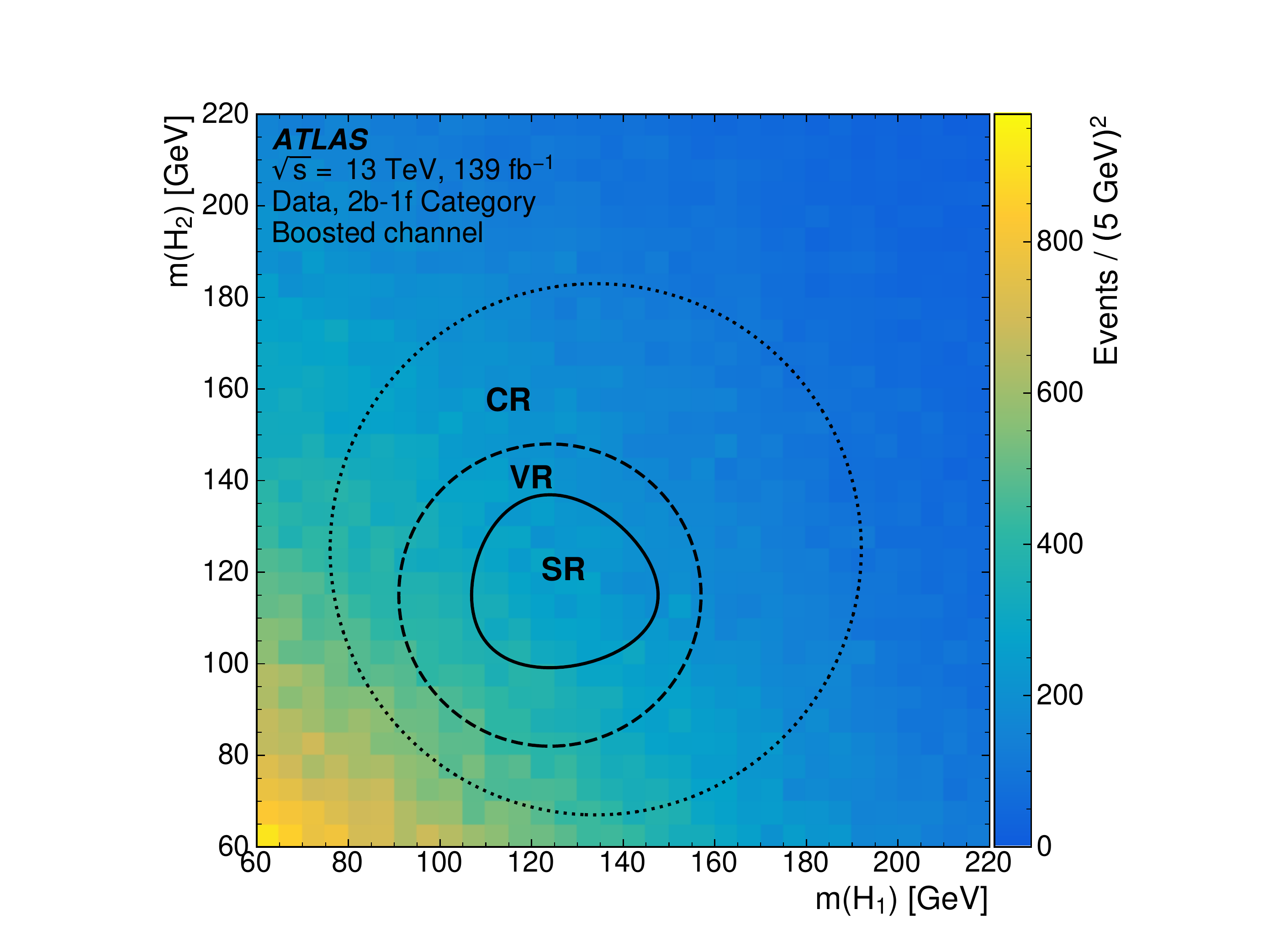}
\caption{
\label{fig:region-def-boosted}
Kinematic region definitions superimposed on the data in the 2\(b\)-1\(f\) category. \(H_1\) and \(H_2\) are the reconstructed Higgs boson candidates, sorted by \(p_\mathrm{T}\).
}
\end{figure}

\noindent
The CR is shifted to higher masses relative to the signal and validation regions in order to maximize the number of selected events while avoiding the low-mass peak of the multijet background distribution.
The definition of these regions in the \(m(H_1)\)--\(m(H_2)\) plane are shown with the 2\(b\)-1\(f\) data in Figure~\ref{fig:region-def-boosted}.
 
In order to ensure orthogonality between the resolved and boosted channels, any events passing the resolved signal region selection are vetoed from the boosted channel. This priority choice results in the best signal sensitivity.
 
\begin{figure}[tbp]
\centering
\subfloat[]{
\label{fig:boosted-acc-scal-cuts}
\includegraphics[width=0.48\textwidth]{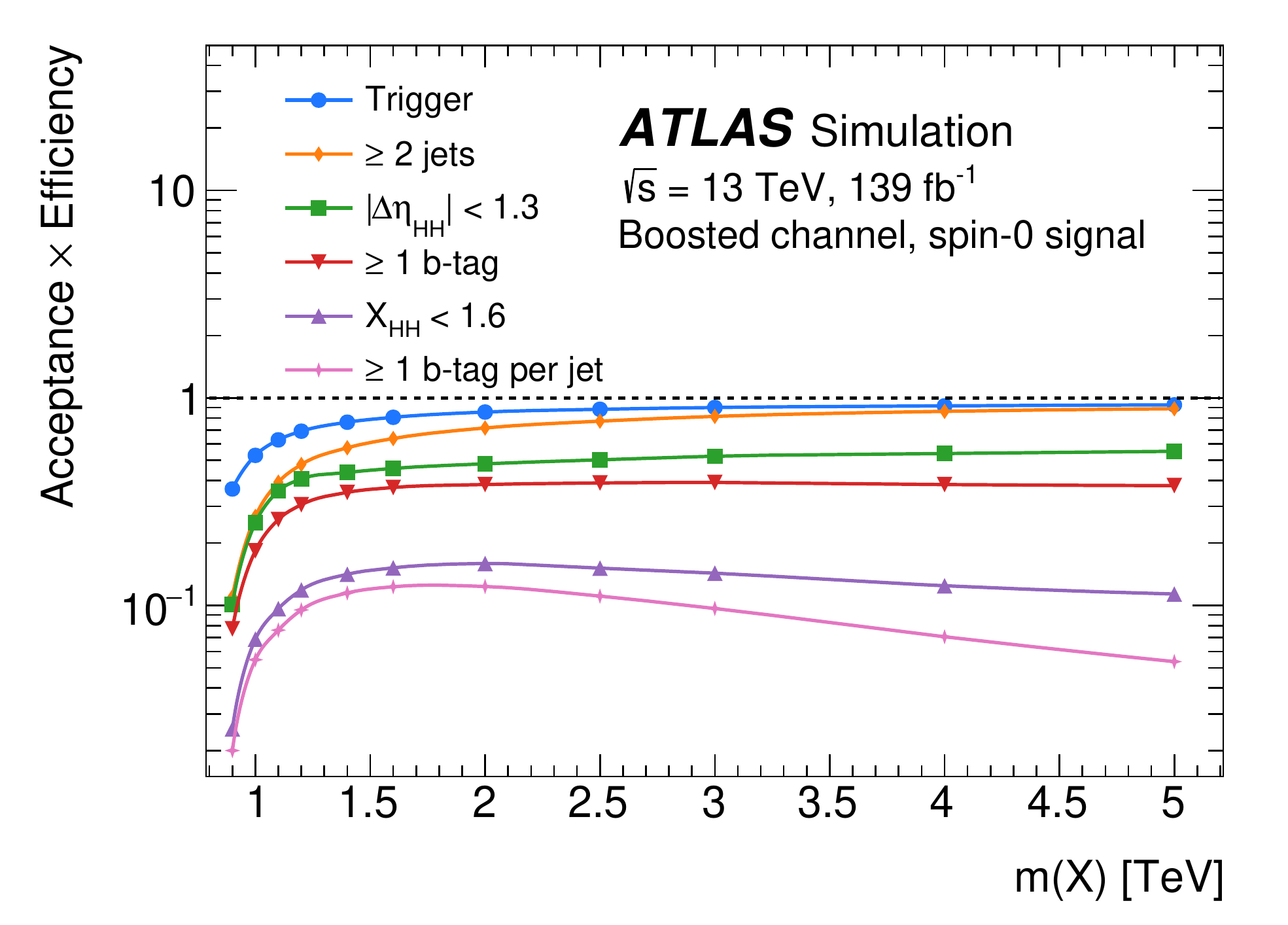}
}
\subfloat[]{
\label{fig:boosted-acc-grav-cuts}
\includegraphics[width=0.48\textwidth]{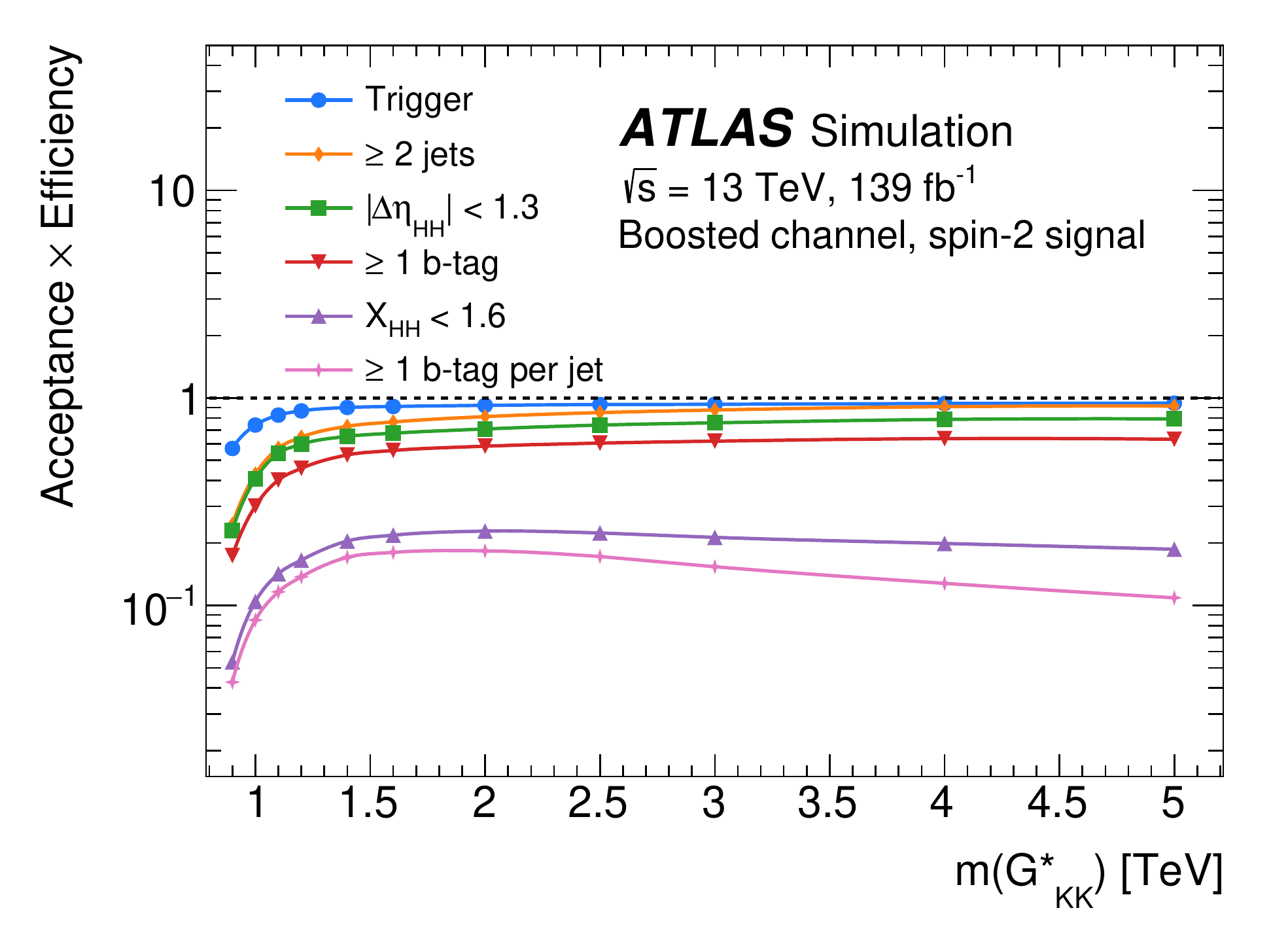}
}
\\
\subfloat[]{
\label{fig:boosted-acc-scal-btag}
\includegraphics[width=0.48\textwidth]{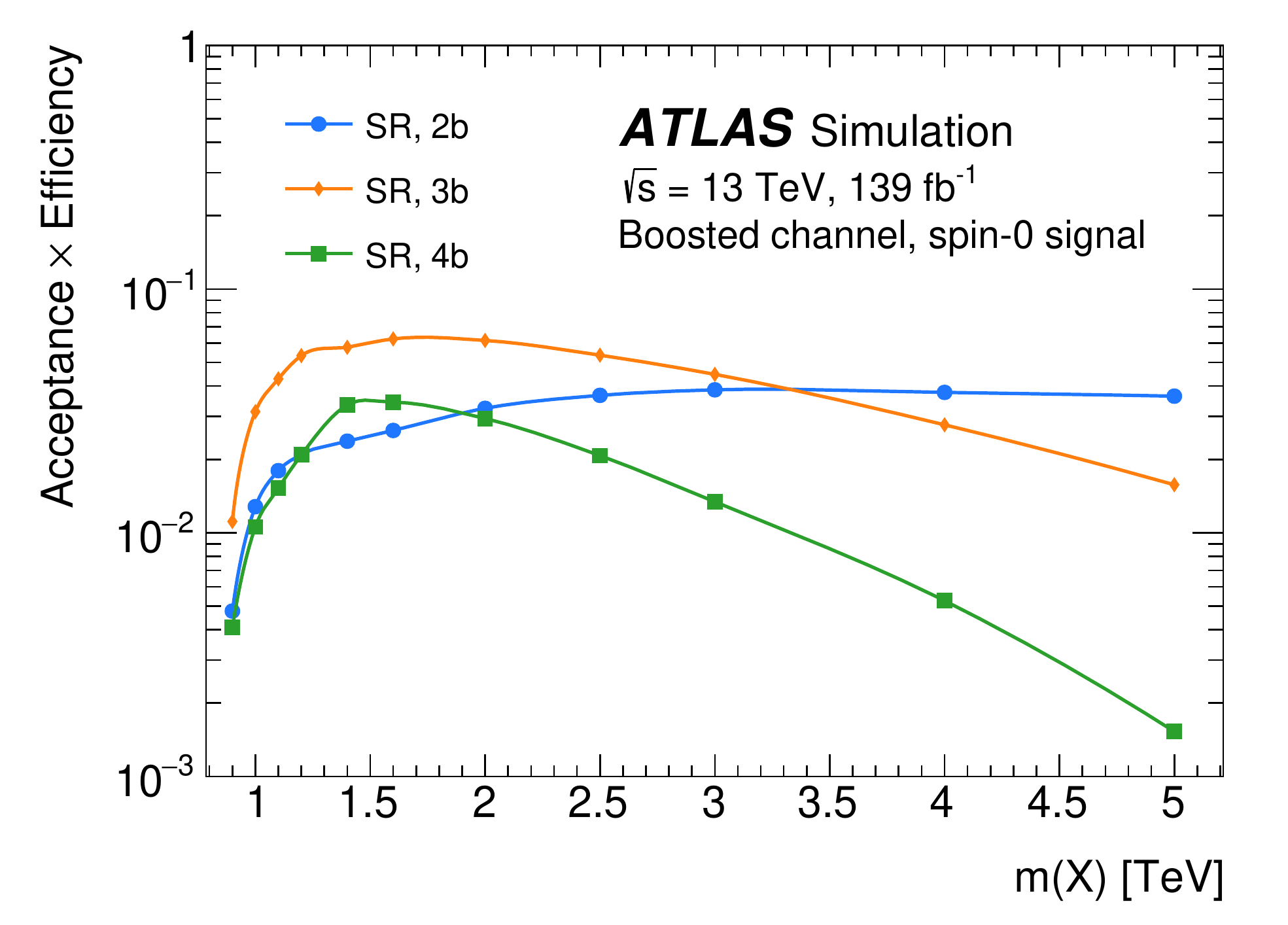}
}
\subfloat[]{
\label{fig:boosted-acc-grav-btag}
\includegraphics[width=0.48\textwidth]{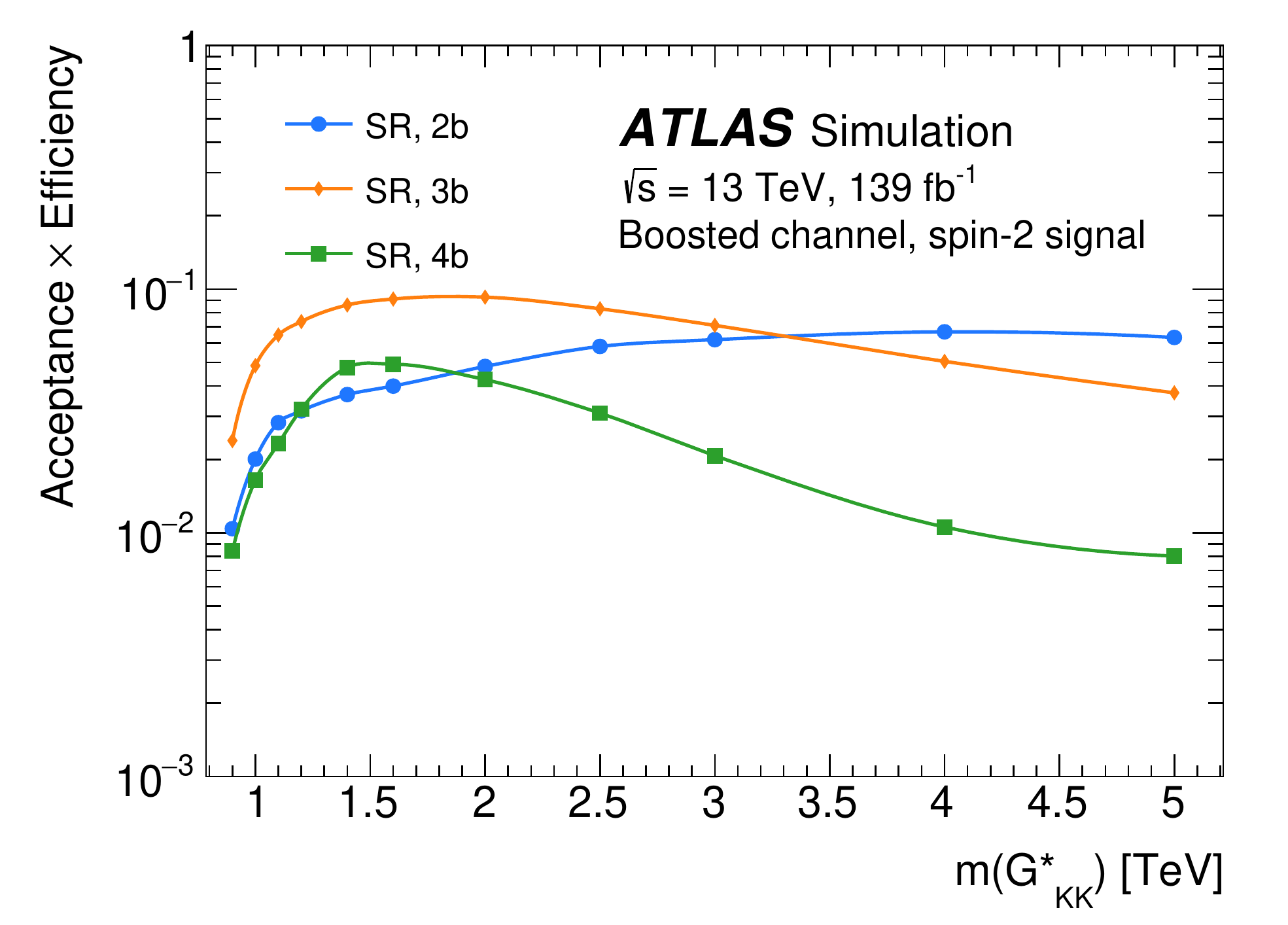}
}
\caption{\label{fig:boosted_acc}
Cumulative signal acceptance times efficiency as a function of the resonance mass for various selection steps in the boosted channel.
The steps up to the \(b\)-tag categorization are shown for
\protect \subref{fig:boosted-acc-scal-cuts} the spin-0 and
\protect \subref{fig:boosted-acc-grav-cuts} the spin-2 signal models.
The efficiencies of the three \(b\)-tag categories are shown for
\protect \subref{fig:boosted-acc-scal-btag} the spin-0 and
\protect \subref{fig:boosted-acc-grav-btag} the spin-2 scenarios;
this efficiency is obtained after the other selection steps including the SR definition.
The signal efficiency in the \(4b\) region has a maximum around \SI{1.5}{\TeV}. Above that value the track-jets start to
merge together, and for the highest resonance masses the \(2b\) category becomes the most efficient.}
\end{figure}

The signal acceptance times efficiency for various steps of the selection is shown in Figure~\ref{fig:boosted_acc}.
 
\FloatBarrier
 
\subsection{Background estimation}
 
As in the resolved channel, the background in the boosted channel is dominated by QCD-induced jet production, which is separated into multijet (light quark) and \ttbar production.
The fractions of \ttbar relative to the total background are 10\%, 15\% and 30\% for the 4\(b\), 3\(b\) and 2\(b\) regions, respectively.
Other background sources, such as single Higgs boson production, SM \(HH\) production, \((Z \to b\bbar)\)+jets, and \(ZZ \to b\bbar b\bbar\) account for \({\leq}1\%\) of the total and are neglected.
 
A data-driven method is used to estimate the multijet background in each of the \(4b\), \(3b\), and \(2b\) signal regions.
The \ttbar background is estimated from MC simulation, with corrections derived from data applied in the 3\(b\) and 2\(b\) regions.
 
The overall normalization of the multijet and \ttbar estimates are obtained from a fit to the CR data in each category.
Two normalization parameters \muqcd and \alphatt per \(b\)-tagging category (here denoted \(n_b\)) are introduced as follows:
 
\begin{equation}\nonumber
N^\text{hi}_{i, \text{data}} = \muqcd^{n_b} (N^\text{lo}_{i, \text{data}} - N^\text{lo}_{i, \ttbar}) + \alphatt^{n_b} N^\text{hi}_{i, \ttbar}.
\end{equation}
 
\begin{figure}[tbp]
\centering
\subfloat[]{
\label{fig:boosted-mh1-2b}
\includegraphics[width=0.32\textwidth]{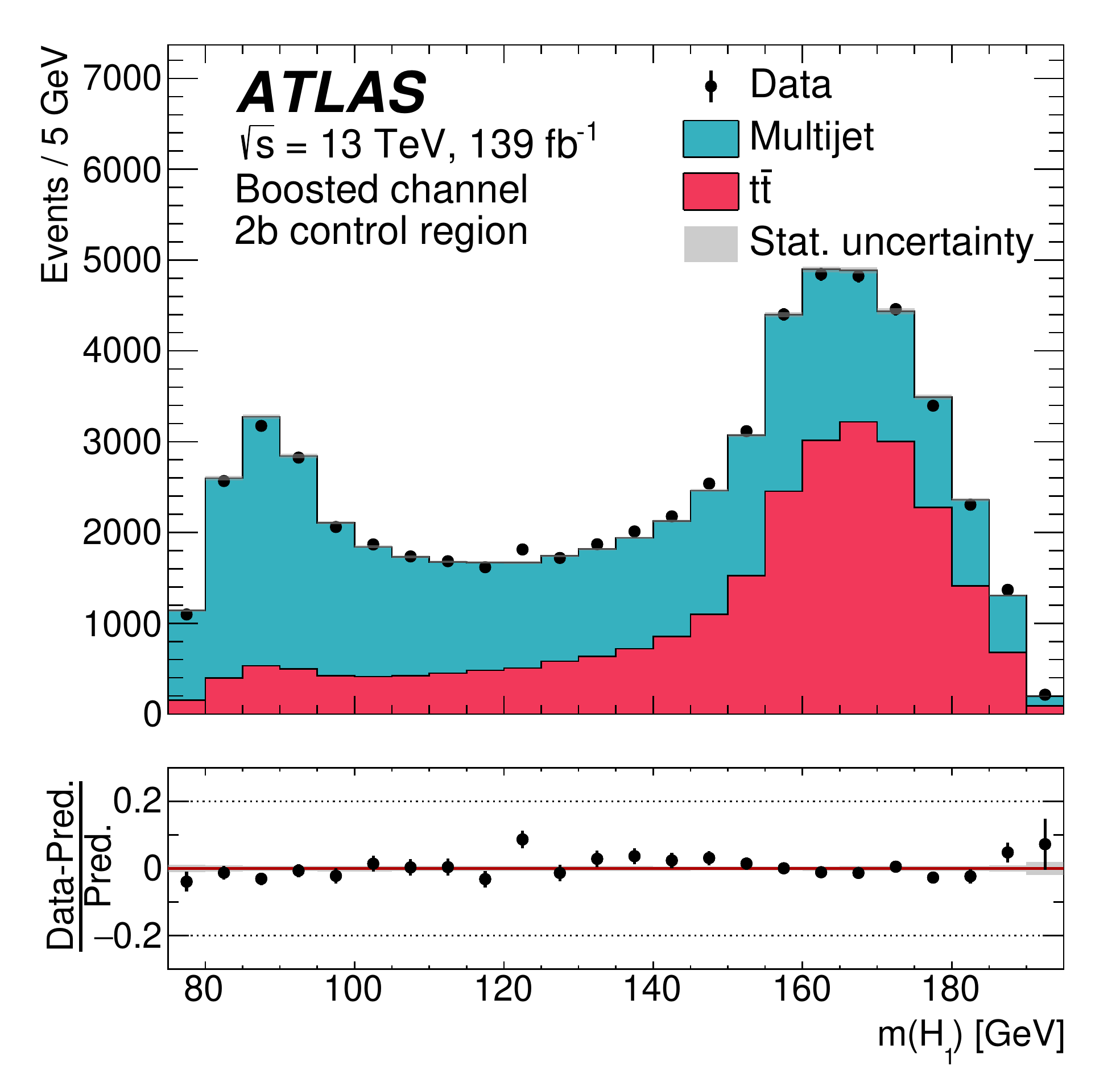}
}
\subfloat[]{
\label{fig:boosted-mh1-3b}
\includegraphics[width=0.32\textwidth]{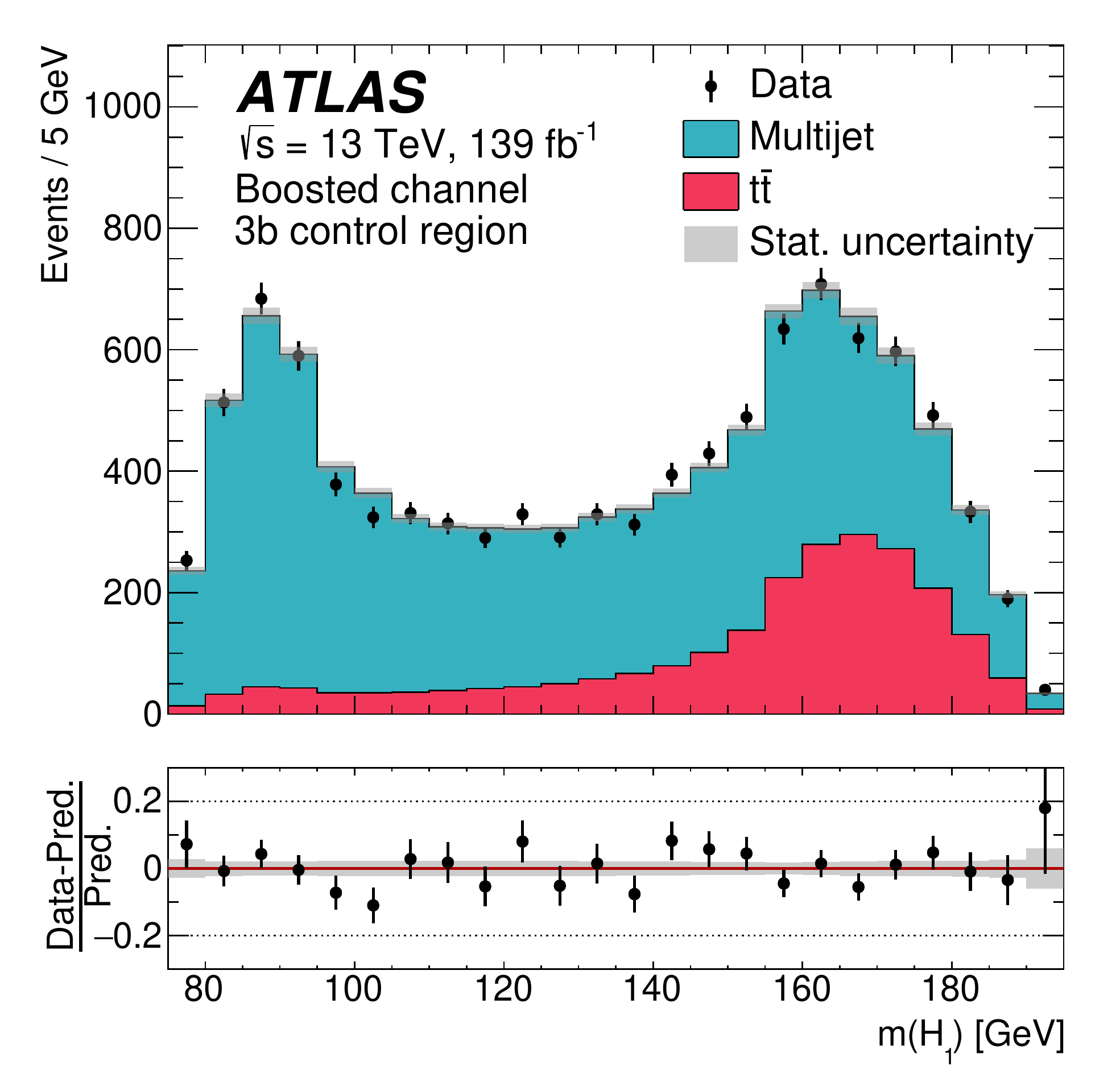}
}
\subfloat[]{
\label{fig:boosted-mh1-4b}
\includegraphics[width=0.32\textwidth]{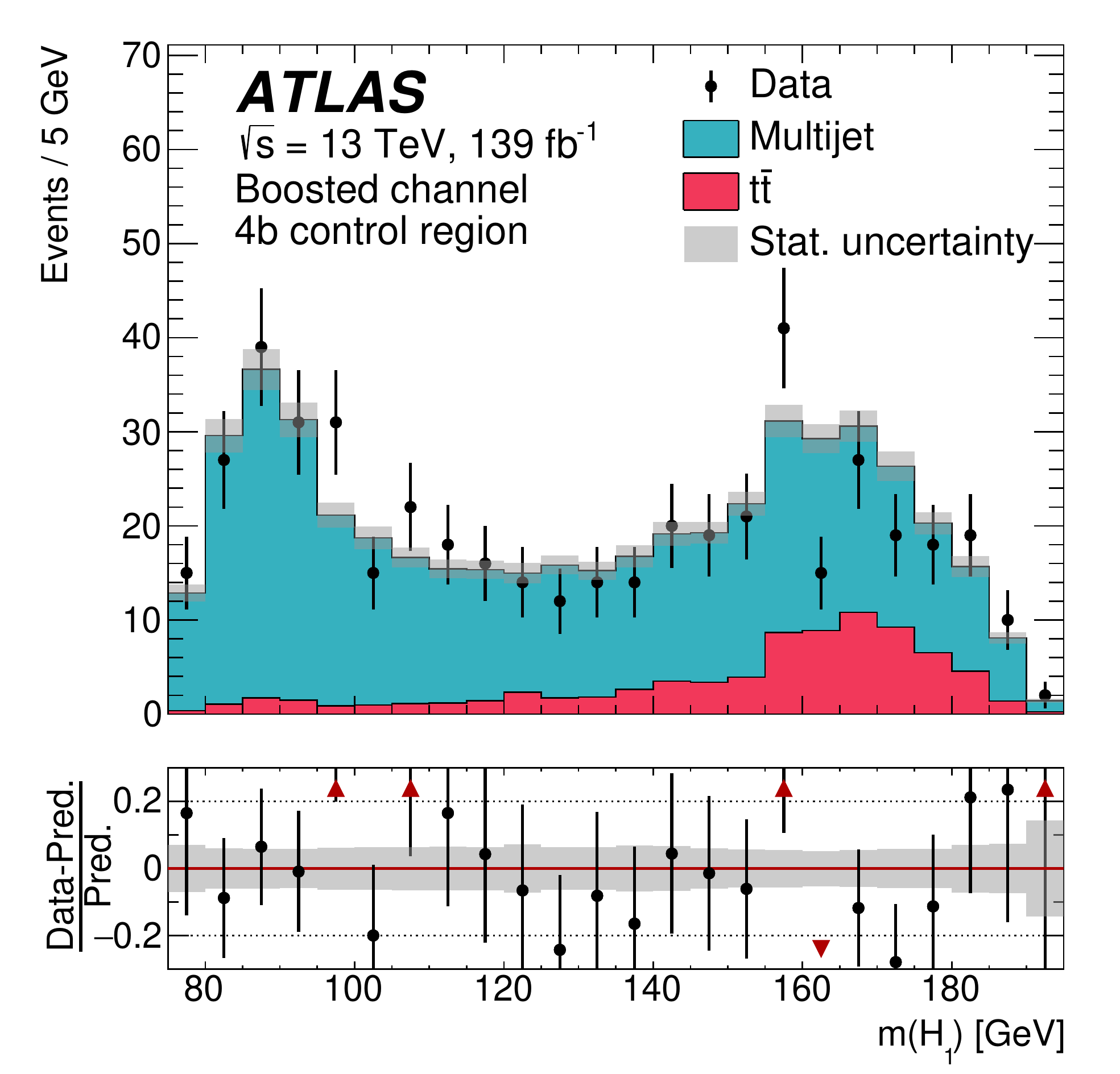}
}
\caption{\label{fig:boosted_normfit}
Reconstructed mass distributions of the leading \(H\) candidate for the data (dots) and the background model (stacked histograms) in the
\protect \subref{fig:boosted-mh1-2b} \(2b\), \protect \subref{fig:boosted-mh1-3b} \(3b\) and \protect \subref{fig:boosted-mh1-4b} \(4b\) control regions.
The error bars on the data points represent the Poisson uncertainties corresponding to their event yields.
The statistical uncertainty of the background model is represented by the gray band.
This distribution is used to normalize the multijet and \ttbar background components.
The enhanced event rates at low and high masses are due to the geometry of the CR.
The bottom panel shows the difference between the data and the background model, normalized to the background model.
}
\end{figure}

\noindent Here, \(N^\text{hi}_i\) and \(N^\text{lo}_i\) denote the number of events in bin \(i\) of the \(H_1\) mass distribution in the high-tag and corresponding low-tag regions, respectively.
The parameter \muqcd scales the multijet background from the low-tag CR to the high-tag CR, and \alphatt corrects the MC estimate in the high-tag region.
The values of \muqcd and \alphatt are determined using a maximum-likelihood fit to the mass of the leading \(H\) candidate after the kinematic reweighting is applied.
The mass of the leading \(H\) candidate discriminates between the two background processes, as shown in Figure~\ref{fig:boosted_normfit}.
The data in the control regions of the \(4b\), \(3b\), and \(2b\) categories are fitted separately, and therefore a total of six floating normalization factors are obtained.
However, the value of \alphatt in the \(4b\) region is fixed to 1, since the fit is insensitive to the \ttbar MC normalization with the available dataset and therefore cannot constrain it.
The results of these normalization fits are summarized in Table~\ref{tab:boosted_norm_results}.
It is assumed that these values of \muqcd and \alphatt are also applicable in the VR and SR; potential deviations from this assumption are accounted for by systematic uncertainties.
The fact that \(\muqcd \ll 1\) implies that the background is much larger in the low-tag categories than in the high-tag categories.
As a result, any potential bias in the high-tag background estimate due to signal contamination in the low-tag categories is much smaller than the signal contribution itself in the high-tag regions.
 
\begin{table}[!ht]
\begin{center}
\caption{Best-fit values for \muqcd and \alphatt, with their statistical uncertainties. The linear correlation coefficient between the two parameters is also given.
The value of \alphatt in the \(4b\) region is fixed to 1, since the data are unable to constrain it significantly.}
 
\begin{tabular}{lS[table-format=1.5,table-number-alignment=center]@{\,$\pm\,$}S[table-format=1.5, table-number-alignment=left]
S[table-format=1.4,table-number-alignment=center]@{\,$\pm\,$}S[table-format=1.4, table-number-alignment=left]
S[table-format=1.4,table-number-alignment=center]@{\,$\pm\,$}S[table-format=1.4, table-number-alignment=left]}
\toprule
Region       & \multicolumn{2}{c}{\(2b\)}    & \multicolumn{2}{c}{\(3b\)}      & \multicolumn{2}{c}{\(4b\)} \\
\midrule
 
\muqcd       & 0.05435   &  0.00056          &  0.1204 & 0.0023                & 0.0272 & 0.0015 \\
\alphatt     & 0.863     &  0.011            &  0.786  & 0.042                 & \multicolumn{2}{c}{1} \\
Correlation  & \multicolumn{2}{c}{\(-0.74\)} &  \multicolumn{2}{c}{\(-0.74\)}  & \multicolumn{2}{c}{0} \\
 
\bottomrule
\end{tabular}
\label{tab:boosted_norm_results}
\end{center}
\end{table}
 
For \(3b\) and \(2b\), a kinematic reweighting procedure is applied to each corresponding low-tag category, analogous to the resolved channel.
For the \(4b\) category, no kinematic reweighting is applied.
This is because the effect of mismodelings due to \(b\)-tagging is small compared to the size of statistical uncertainties in this category.
Instead of an NN for constructing the reweighting function, an iterative spline method based on the one used in Ref.~\cite{EXOT-2016-31} is implemented here.
 
The difference between these low-tag and high-tag regions is that the low-tag events have an untagged \(H\) candidate (no \(b\)-tagged track-jets),
while high-tag events instead have a tagged \(H\) candidate (exactly one \(b\)-tagged track-jet, since only the \(3b\) and \(2b\) categories are considered here).
Therefore, the reweighting applied to low-tag events is based on their untagged \(H\) candidates, with the aim of matching the kinematics of the tagged \(H\) candidates in high-tag events.
This reweighting is derived purely in the low-tag regions; the tagged \(H\) candidates in the 1\(b\)-1\(f\) category are used to define the target.
 
The following kinematic distributions are used to construct the reweighting, for which leading and subleading refer to an ordering in \pt:
 
\begin{enumerate}
\item \pt of the \(H\) candidate,
\item \pt of the chosen track-jet,
\item \(\eta\) of the chosen track-jet, and
\item \(\Delta R\) between the leading and subleading track-jets (for \(H\) candidates with at least two track-jets).
\end{enumerate}
 
\noindent The ``chosen'' track-jet is the \(b\)-tagged one for tagged \(H\) candidates and a random one for untagged \(H\) candidates.
In tagged \(H\) candidates with two track-jets, the leading and subleading track-jets have roughly equal probabilities to be the \(b\)-tagged one, so this random selection does not introduce a significant bias.
Separate distributions are constructed for leading and subleading \(H\) candidates, as well as for leading and subleading track-jets.
 
At each iteration \(i\), cubic splines are fitted to the ratios of tagged to untagged distributions, and the weights are updated according to
 
\begin{equation}\nonumber
w_i(\vec{x}) = w_{i-1}(\vec{x}) \times \left[\left(\prod_j f_{ij}(x_j) - 1\right) \times r_i + 1\right],
\end{equation}
 
\noindent where \(j\) indexes the different reweighting variables \(x_j\), \(f_{ij}\) denotes the spline functions,
and the ``learning rate'' \(r_i\) controls how much the weight can change with each iteration.
This is set to \(r_i = 1 - 0.5^i\), and ten iterations are used, after which convergence is observed.
Suppressing the learning rate for early iterations is intended to avoid instabilities.
This reweighting function is applied to the low-tag data sample which contains multijet and \ttbar events.
In order to obtain the multijet model, the \ttbar contribution is subtracted from data, and for that purpose the \ttbar events in the low-tag regions are therefore also reweighted.
The \ttbar distributions in the high-tag regions are not reweighted.
 
\begin{figure}[tbp]
\centering
\subfloat[]{
\label{fig:boostedSmooth-2b}
\includegraphics[width=0.32\textwidth]{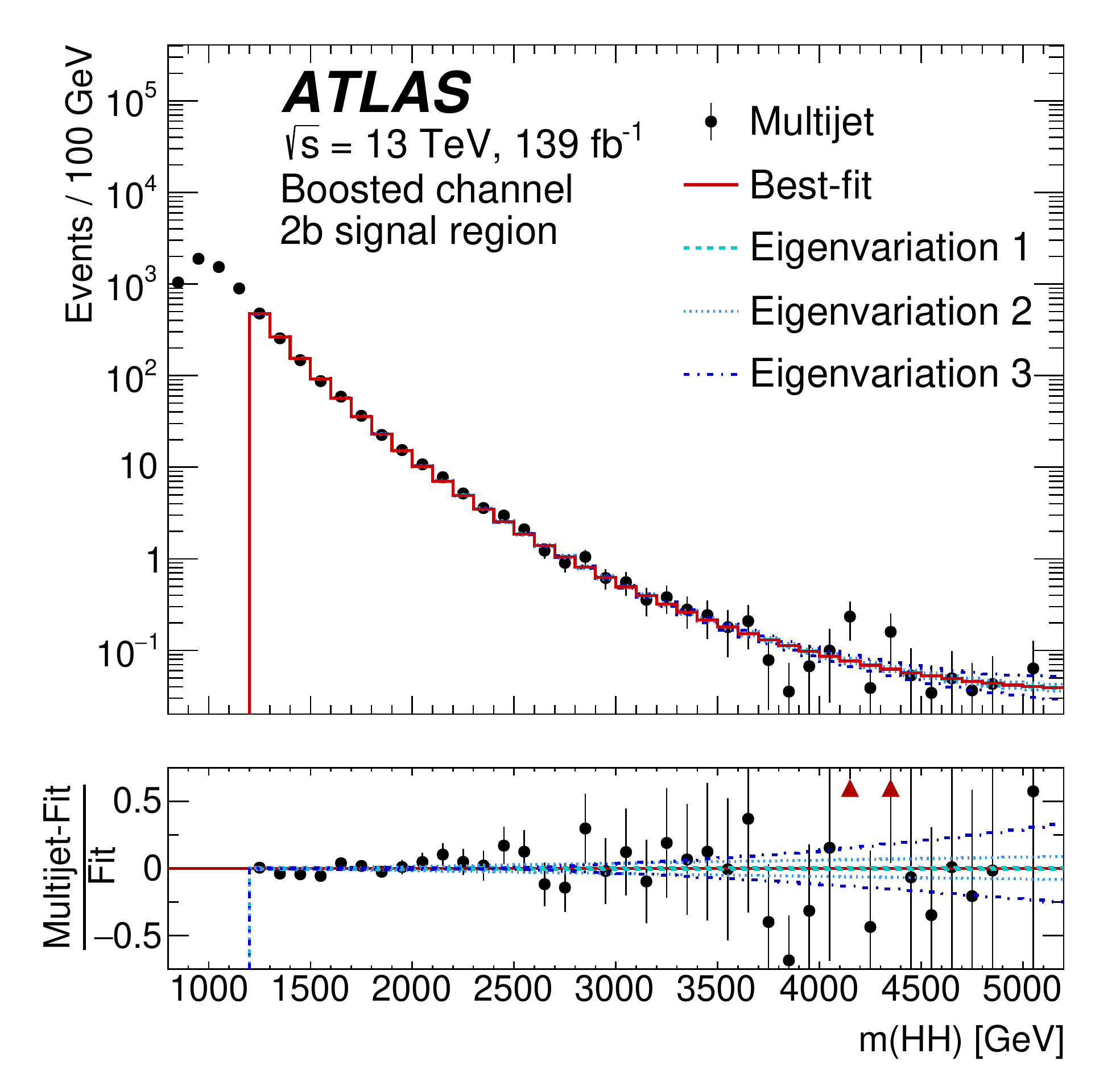}
}
\subfloat[]{
\label{fig:boostedSmooth-3b}
\includegraphics[width=0.32\textwidth]{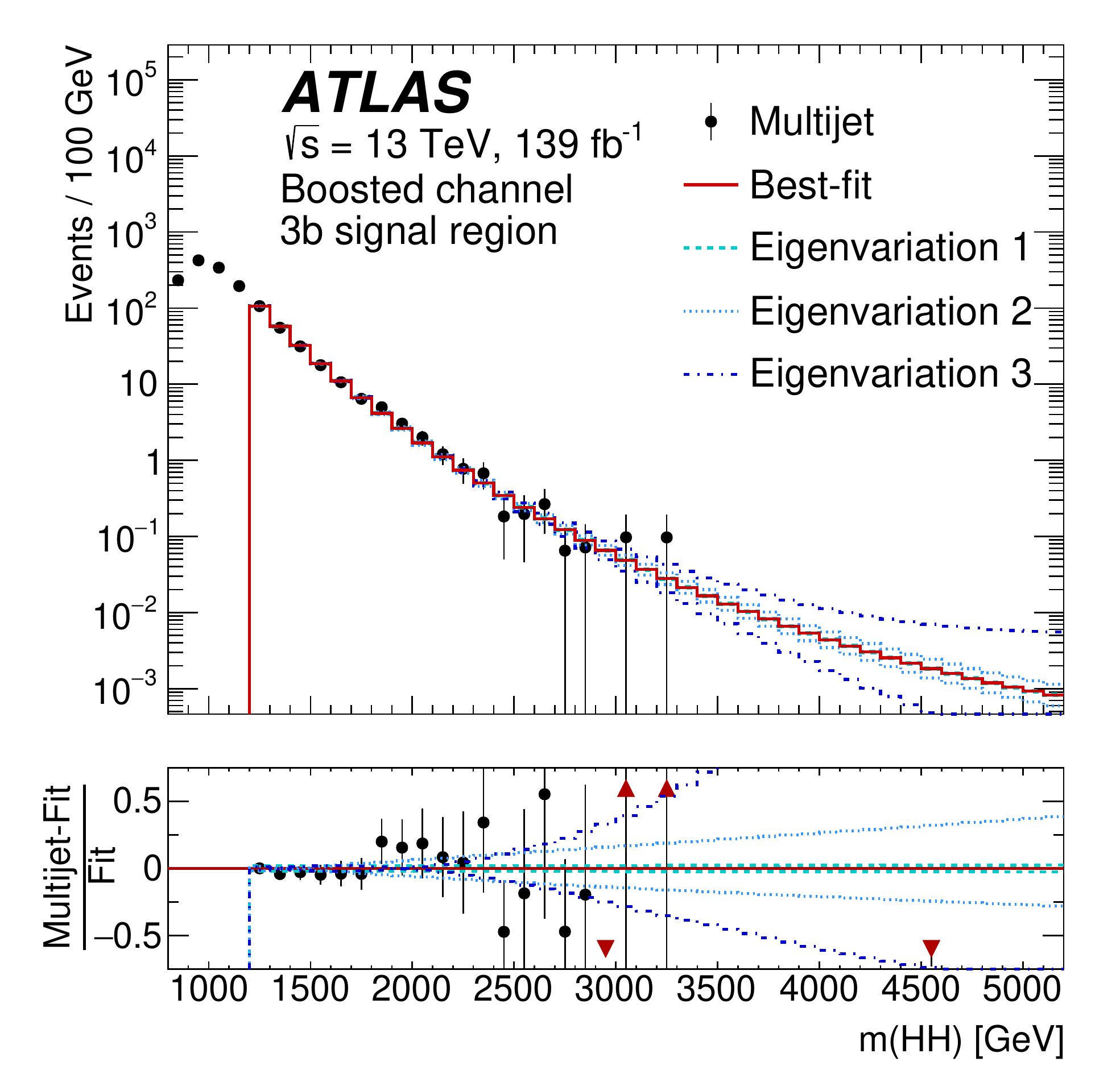}
}
\subfloat[]{
\label{fig:boostedSmooth-4b}
\includegraphics[width=0.32\textwidth]{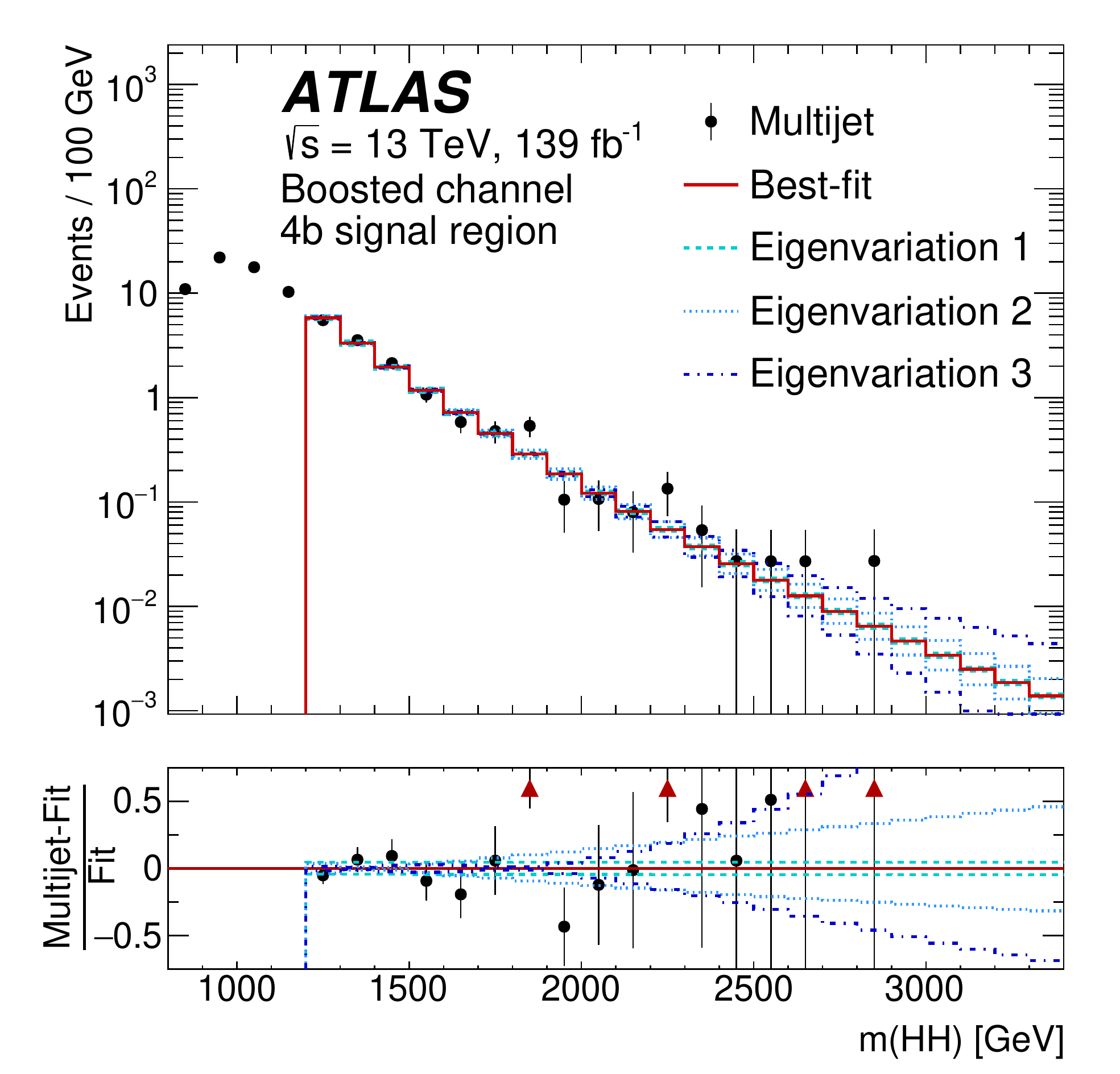}
}
\caption{\label{fig:boostedSmooth}
Smoothing fit (solid line) applied to the \(HH\) invariant mass spectrum of the multijet background estimate (dots with error bars) in the
\protect \subref{fig:boostedSmooth-2b} \(2b\), \protect \subref{fig:boostedSmooth-3b} \(3b\), and \protect \subref{fig:boostedSmooth-4b} \(4b\) signal regions.
The effects of the fit function parameter eigenvariations (dashed and dashed-dotted lines) are also shown, indicating the effective statistical uncertainties.
The bottom panel shows the difference between the unsmoothed and smoothed multijet background estimates, normalized to the smoothed estimate.
}
\end{figure}

In order to reduce the effect of statistical fluctuations in the background at high \mhh, the following function is fitted to the reweighted multijet and \ttbar distributions for \(\mhh \geq \SI{1200}{\GeV}\):
 
\begin{equation}
\label{eq:boosted-mhh-fit}
f(x) = \frac{e^{-p_0}}{x^2}(1-x)^{p_1 - p_2 \ln x},
\end{equation}
 
\noindent where the \(p_i\) are dimensionless free parameters and \(x \equiv \mhh/\sqrt{s}\).
This function and similar ones have been used to fit falling dijet and multijet spectra in similar analyses (e.g. Refs.~\cite{EXOT-2016-31,EXOT-2014-02}).
The results of these smoothing fits for the multijet background model are shown in Figure~\ref{fig:boostedSmooth}.
Due to the small number of events in the \(4b\) category, the shape of the \(t\bar{t}\) distribution in this region is taken from the \(3b\) category and scaled to the yield in the \(4b\) category.
This is found to be consistent with the shape in the \(4b\) category within statistical uncertainties.
 
\begin{figure}[tbp]
\centering
\subfloat[]{
\label{fig:boosted-mhh-vr-2b}
\includegraphics[width=0.32\textwidth]{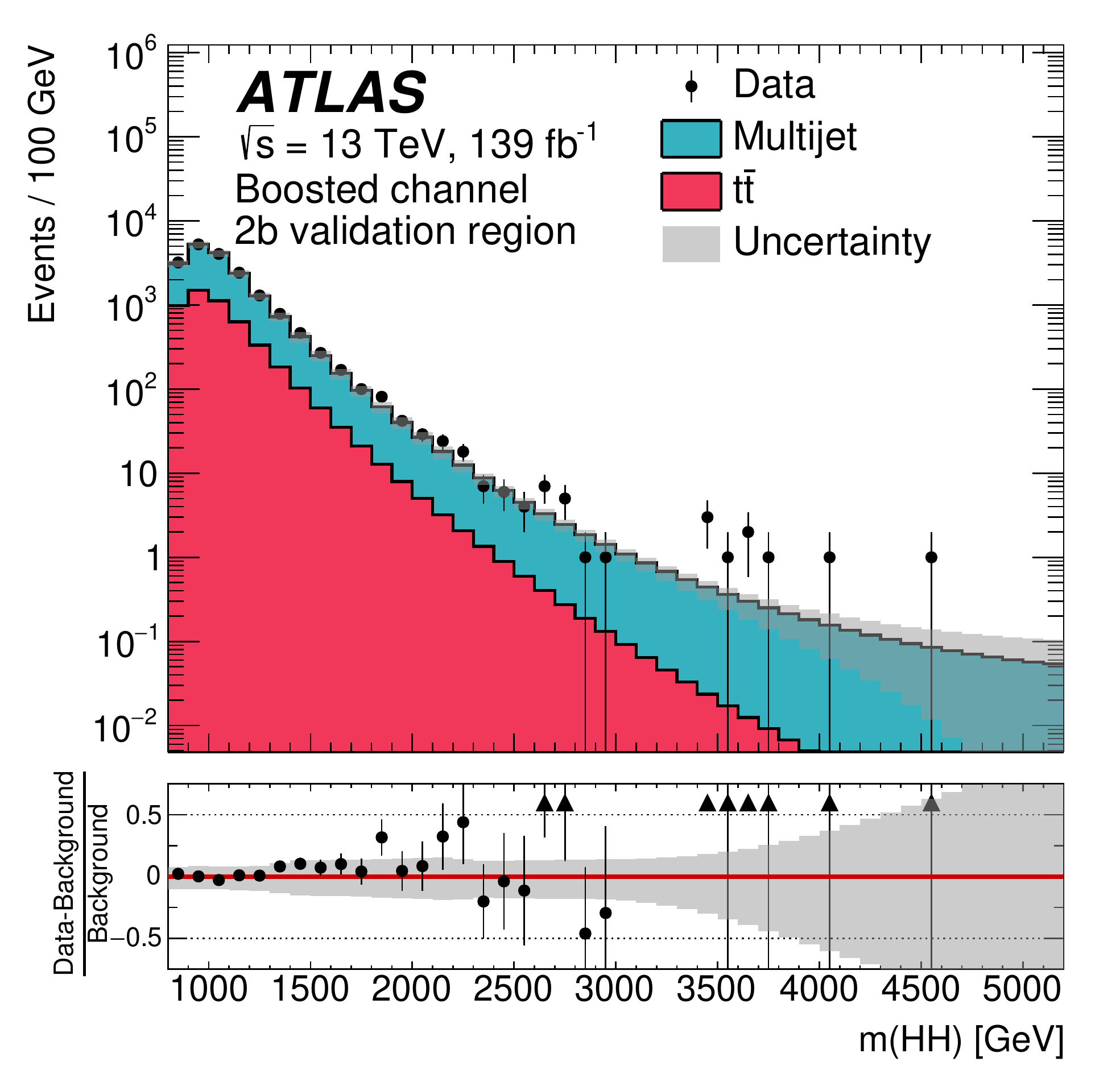}
}
\subfloat[]{
\label{fig:boosted-mhh-vr-3b}
\includegraphics[width=0.32\textwidth]{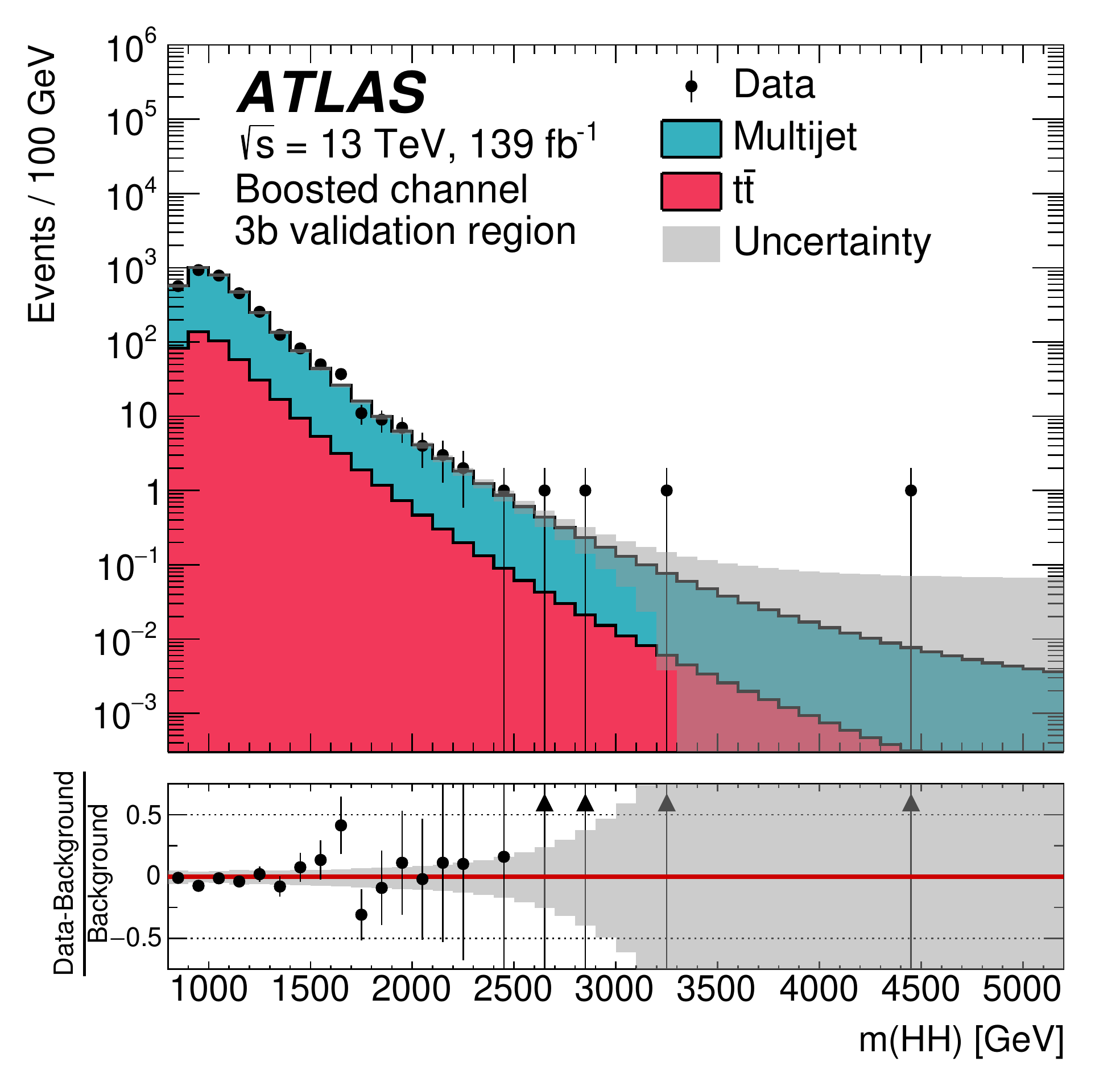}
}
\subfloat[]{
\label{fig:boosted-mhh-vr-4b}
\includegraphics[width=0.32\textwidth]{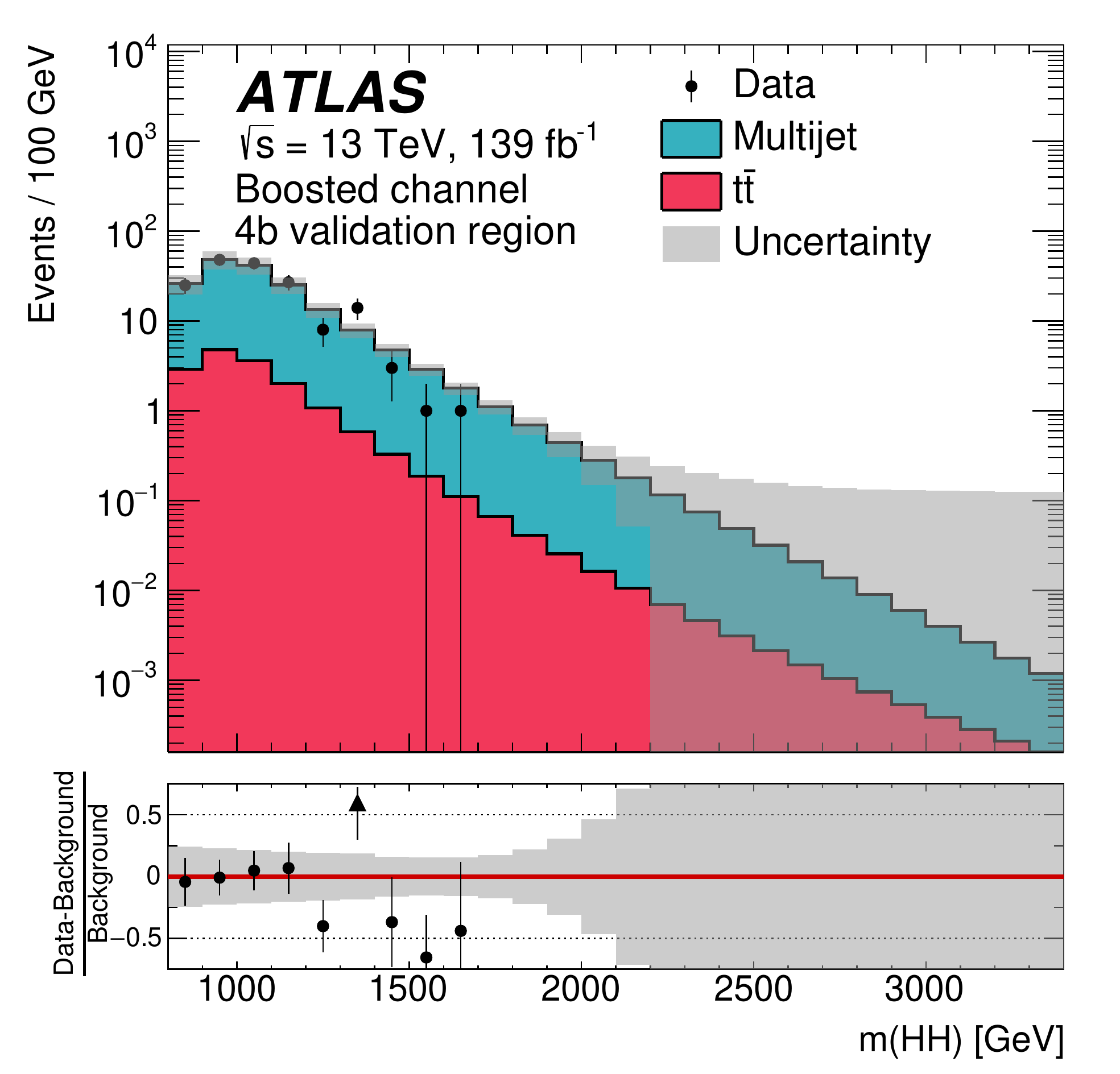}
}
\caption{\label{fig:boosted-mhh-vr}
Comparison of the background model (stacked histograms) with data (dots) in the
\protect \subref{fig:boosted-mhh-vr-2b} \(2b\), \protect \subref{fig:boosted-mhh-vr-3b} \(3b\), and \protect \subref{fig:boosted-mhh-vr-4b} \(4b\) validation regions.
The error bars on the data points represent the Poisson uncertainties corresponding to their event yields.
The background uncertainty (gray band) is computed by adding all individual components in quadrature and is not allowed to extend below zero.
The bottom panel shows the difference between the data and the background model, normalized to the background model.
}
\end{figure}

The background model outside the region where it is derived is checked using the VR, as shown in Figure~\ref{fig:boosted-mhh-vr}.
Good agreement between the background model and the data in the VR is observed; residual differences between the CR and VR are used to estimate a systematic uncertainty as described in Section~\ref{sec:boostedSyst}.
 
\subsection{Systematic uncertainties}
\label{sec:boostedSyst}
 
The boosted channel is generally limited by statistical uncertainties, especially for higher resonance masses.
Systematic uncertainties in the background and hypothesized signals are nonetheless fully accounted for and described here.
 
The uncertainty in the data-driven background estimate is split into three normalization components and five shape components.
Each of these components is evaluated separately for the \ttbar background (from simulation) and the multijet background.
 
The first normalization component is associated with the assumption that the normalization factor \muqcd derived in the CR is also applicable in the SR.
To evaluate this uncertainty, a Gaussian process technique~\cite{GaussianProcess} is used to interpolate the multijet contribution in the SR.
First, a fit is performed on the two-dimensional \(m(H_1)\)--\(m(H_2)\) distribution, with the SR data removed, to determine the parameters for a Gaussian two-point correlation function in each of the low-tag and high-tag regions.
The correlation scales are optimized as part of the fit and are found to be >\SI{100}{\GeV} in all cases, so the interpolation is able to smoothly fit across the SR.
Second, the resulting correlation function is used to construct an estimate for all points in the \(m(H_1)\)--\(m(H_2)\) plane.
Event yields in each region, in particular the SR, are determined from this result.
The uncertainty is finally defined as the difference of high-tag to low-tag yield ratio between the SR and the CR.
This is applied only to the multijet component of the background; the \ttbar estimate from MC is subtracted from the data when deriving it.
 
The second normalization component is associated with the choice of control region.
The suitability of this choice is validated by comparing the background model with data in signal-depleted regions; however, a small dependence of the background estimate on this choice remains.
This is accounted for by defining slightly different CRs and re-deriving the background model from each of them.
The largest difference between any of these and the nominal background model is taken as a normalization uncertainty.
The alternative CRs are defined by raising or lowering the values of \(m(H_1)\) and \(m(H_2)\) by \SI{3}{\GeV} (independently, for four variations).
Additional variations are defined by increasing or decreasing the value of the \(\Rhh^{\mathrm{CR}}\) requirement by \SI{3}{\GeV} while applying the opposite change to the \(\Rhh^{\mathrm{VR}}\) requirement,
effectively making the CR thicker or thinner in the \(m(H_1)\)--\(m(H_2)\) plane.
In all cases, a veto on the SR is maintained.
This procedure results in uncertainties of 0.9\%, 1.6\%, and 6.0\% for the \(2b\), \(3b\), and \(4b\) regions, respectively.
 
The third component of the background normalization uncertainty is associated with the statistical uncertainty of the normalization fit.
The best-fit values of \muqcd and \alphatt are varied along the eigenvectors of their covariance matrix.
The varied values are propagated through the background estimation procedure to evaluate the resulting effect.
Since \alphatt is fixed to a value of 1 for the \(4b\) category, only one variation is performed in that case, corresponding to \muqcd.
This variation changes the normalizations of the individual background components, resulting in different total background shapes.
 
The first component of the background shape uncertainty comes from the limited sample size used to fit the smoothing function in Eq.~(\ref{eq:boosted-mhh-fit}).
This is taken into account by an eigenvariation method using the covariance matrix of the function parameters.
Function variations are defined by varying the best-fit parameters according to the eigenvectors, scaled to the square root of the corresponding eigenvalues.
These varied functions are then treated as components of the uncertainty in the background shape.
Three variations are used for each of the multijet and \ttbar background components.
 
Two further components of the shape uncertainty in the background model are associated with choosing the fit function that is used in Eq.~(\ref{eq:boosted-mhh-fit}).
As the function and fit range are arbitrary and chosen based only on empirical results from the CR and VR, both are varied.
Seven alternative functional forms are used, based on the scheme from Ref.~\cite{EXOT-2014-02}.
The nominal fit is compared with the other seven, and those which differ the most
from the nominal one in each direction are used to define an envelope for the shape uncertainty.
The result of this procedure is shown in Figure~\ref{fig:boostedSmooth}.
The fit range is also varied by changing its upper and lower bounds by \SI{100}{\GeV} in each direction.
Again, the results which differ the most from the nominal one are used to construct an envelope, defining another shape uncertainty.
In cases where all variations fall on one side of the nominal result, an envelope is constructed by symmetrization.
 
The fourth component of the shape uncertainty is the ``residual'' one, associated with shape differences between the CR and SR.
This is associated only with the multijet background, as direct simulation of the SR is used for the \ttbar background shape.
This uncertainty is evaluated by comparing the multijet background model with the data in the VR, after subtracting the \ttbar component determined from simulation.
Since the events in the VR are kinematically closer to the SR than events in the CR are, this comparison of the background models covers residual mismodeling connected to the extrapolation over the regions.
The multijet models agree within statistical uncertainties for the \(3b\) and \(4b\) regions, so this uncertainty is only derived for \(2b\) events.
The shape of the uncertainty is defined as the ratio of the multijet background model to the \ttbar-subtracted VR data.
Empty bins and bins with relative statistical uncertainty over 50\% are discarded
to suppress unphysical effects from statistical fluctuations, and the remainder are smoothed and symmetrized to obtain the final shape.
 
The fifth component of the shape uncertainty is associated with the nonclosure of the background estimation method itself.
This is evaluated by running the full background estimation procedure on multijet MC simulation.
The reweighting is derived in the CR and applied to the low-tag events in both the VR and SR, which are summed into a single region for the purpose of evaluating this uncertainty.
The ratio of the resulting estimate to the actual MC prediction in that summed region is determined as a function of \mhh.
A linear function is fitted to this set of ratios in each high-tag category, in order to mitigate the effects of statistical fluctuations.
The result is then symmetrized to form an envelope, which defines this component of the shape uncertainty.
It is applied only to the multijet component of the background.
For \(2b\) and \(3b\) categories, the relative size of this uncertainty is 5\% or less for \(\mhh < \SI{2}{\TeV}\), ranging up to about 15\% for the largest values of \mhh.
For the \(4b\) category, the relative size of this uncertainty is less than 15\% over the whole \mhh range.
 
Several experimental uncertainties are considered.
These affect only the signal and \ttbar background models directly, as the multijet background is estimated from data alone.
However, they have indirect effects on the multijet background model due to the \ttbar subtraction applied in the data-driven procedure.
These are accounted for by propagating the corresponding variations through the background estimation procedure.
Uncertainties in the jet energy and mass scales and resolutions are treated according to the prescriptions in Refs.~\cite{JETM-2018-02,ATLAS-CONF-2017-063}.
Uncertainties in the \(b\)-tagging efficiency are treated according to the prescription in Ref.~\cite{FTAG-2018-01}; this follows exactly the same procedure as in the resolved channel.
Unlike in the resolved channel, no trigger efficiency uncertainties are required: no \(b\)-tagging is done at trigger level
and the jet \pt requirement is high enough that the triggers are fully efficient.
As in the resolved channel, the uncertainty in the total integrated luminosity is 1.7\%~\cite{ATLAS-CONF-2019-021},
obtained using the LUCID-2 detector~\cite{Avoni:2018iuv}.
 
Theory uncertainties in the \ttbar background model are evaluated for the matrix element and parton shower parts of the calculation.
Matrix element uncertainties are computed by raising and lowering the factorization and renormalization scales by a factor of two.
The \POWHEG damping parameter \hdamp is also varied upwards by a factor of two.
Additionally, a comparison with an alternative matrix element generator, \MGNLO[2.6.0], is performed and the difference is taken as an uncertainty component.
Uncertainties in the initial-state and final-state radiation modeling are evaluated using both scale variations and eigenvariations of the A14 tune~\cite{ATL-PHYS-PUB-2014-021}.
Parton shower uncertainties are obtained by generating alternative samples which are showered using \HERWIG[7.1.3] instead of \PYTHIA[8.230] and taking the difference between the resulting \mhh distributions.
The effects of PDF uncertainties are evaluated by comparing the nominal \mhh distribution with those obtained from a set of 100 weight variations, but are found to be much smaller than the statistical uncertainty and are therefore neglected.
 
Theory uncertainties in the signal models are evaluated using exactly the same method as for the resolved channel.
Here, the parton shower and underlying event are found to have a 10\% effect on the signal acceptance, independently of the resonance mass.
Uncertainties in the matrix element calculation are again found to be negligible, and are not included in the statistical interpretation.
 
\subsection{Results}
 
\begin{figure}[tbp]
\centering
\subfloat[]{
\label{fig:boosted-mhh-2b}
\includegraphics[width=0.45\textwidth]{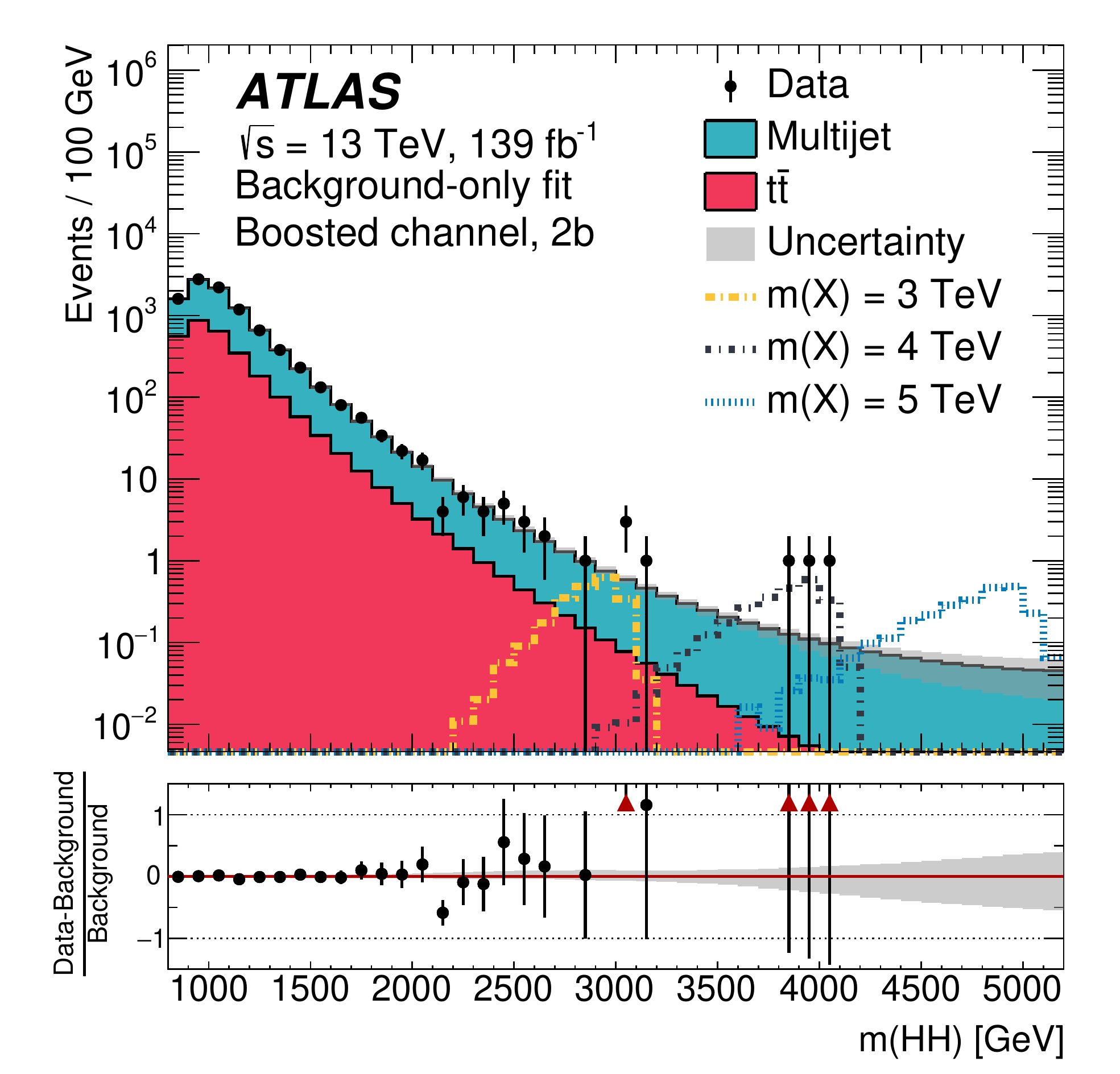}
}
\subfloat[]{
\label{fig:boosted-mhh-3b}
\includegraphics[width=0.45\textwidth]{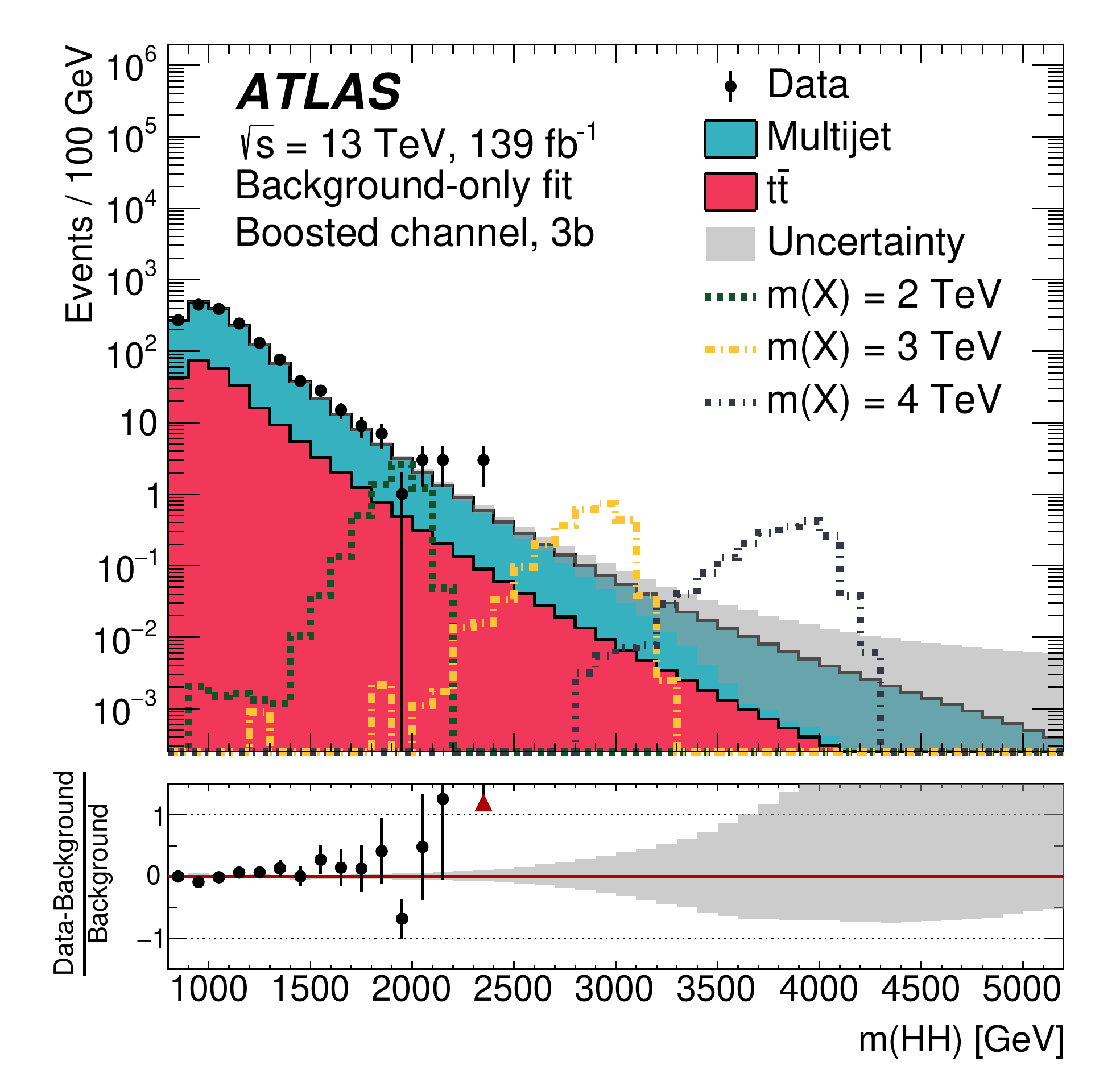}
}\\
\subfloat[]{
\label{fig:boosted-mhh-4b}
\includegraphics[width=0.45\textwidth]{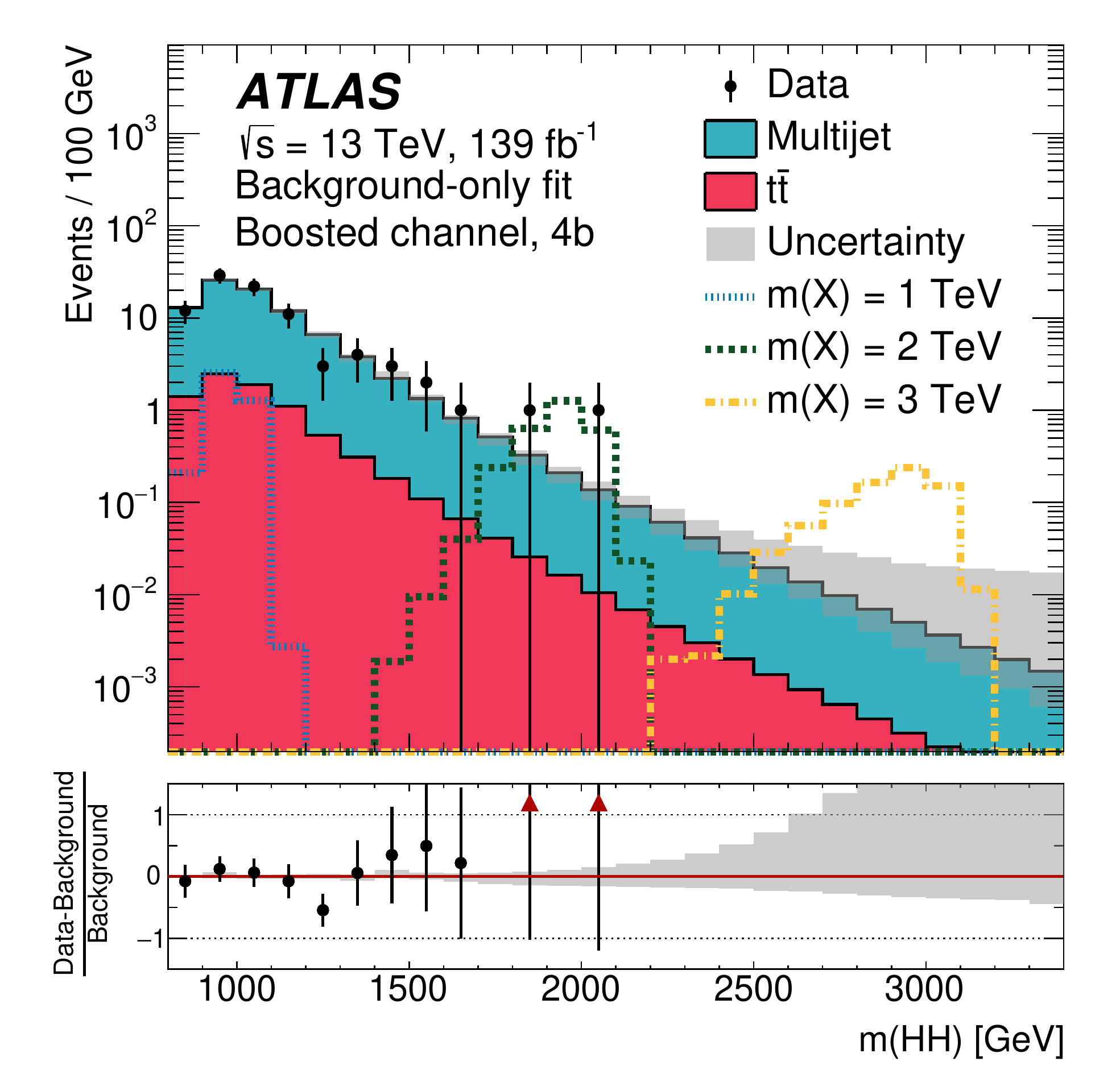}
}
\caption{\label{fig:boosted-mhh}
The \mhh distributions in the boosted \protect \subref{fig:boosted-mhh-2b} \(2b\),
\protect \subref{fig:boosted-mhh-3b} \(3b\), and
\protect \subref{fig:boosted-mhh-4b} \(4b\) signal regions (dots), after the fit under the background-only hypothesis.
The error bars on the data points represent the Poisson uncertainties corresponding to their event yields.
The background model (stacked histogram) is shown with its total post-fit uncertainty (gray band).
The uncertainty bands are defined using an ensemble of curves constructed by sampling a multivariate Gaussian probability density function built from the covariance matrix of the fit.
Representative spin-0 signal hypotheses (dashed and dashed-dotted lines) are overlaid, normalized to the overall expected limits on their cross-sections.
The bottom panel shows the difference between the data and the background model, normalized to the background model.
No significant excess of data relative to the SM background is observed.
}
\end{figure}

The \mhh distributions for data and the estimated background are shown in Figure~\ref{fig:boosted-mhh} for the three categories.
The data agree well with the background and no significant excess is observed.
The numbers of events for data, background, and several signal hypotheses are presented in Tables~\ref{tab:boosted-yields-spin0} and \ref{tab:boosted-yields-spin2}.
These event yields are integrated over a set of \mhh bins containing approximately 90\% of the signal in each case.
These windows are defined such that their first and last bins contain the \(5^\text{th}\) and \(95^\text{th}\) percentile of the distribution, respectively.
This range is larger for the spin-2 signal because the benchmark model predicts a wider resonance.
The statistical interpretation of the data is discussed in Section~\ref{sec:results}.
 
\begin{table}[tbp]
\begin{center}
\caption{Boosted signal region data, estimated background, and signal event yields in \mhh windows containing roughly 90\% of each signal, for representative spin-0 mass hypotheses.
The signal is normalized to the overall expected limit on its cross-section; its uncertainties are evaluated by adding all individual components in quadrature.
The background yields and uncertainties are evaluated after a background-only fit to the data.
The \(4b\) category is not used for \(m(X) > \SI{3}{\TeV}\) and the \(2b\) category is not used for \(m(X) < \SI{2}{\TeV}\).
}
\label{tab:boosted-yields-spin0}
\begin{tabular}{lcc S[table-format=4.0] S[table-format=4.3]@{\(\;\pm\;\)}S[table-format=2.3] S[table-format=4.2]@{\(\;\pm\;\)}S[table-format=1.4]}
\toprule
\(m(X)\) [\GeV]        & Category &  \mhh range [GeV] & \multicolumn{1}{c}{Data} & \multicolumn{2}{c}{Background model} & \multicolumn{2}{c}{Spin-0 signal model} \\
\midrule
\multirow{2}{*}{1000} & \(3b\) &  ~~[900, 1200] & 1076 &  1109 & 28   & 11.2 & 2.0 \\
& \(4b\) &  ~~[900, 1200] &   62 &  58.4 & 5.8  & 3.85 & 0.67 \\
\cline{1-8}
\multirow{3}{*}{2000} & \(2b\) & [1800, 2200] &  77 &  77.8 & 3.5   & 2.69  & 0.92 \\
& \(3b\) & [1800, 2200] &  14 & 11.47 & 0.76  & 5.2  & 1.6 \\
& \(4b\) & [1800, 2200] &   2 &  0.76 & 0.11  & 2.5  & 1.1 \\
\cline{1-8}
\multirow{3}{*}{3000} & \(2b\) & [2600, 3200] &   7 &  5.78 & 0.45  & 2.01  & 0.84 \\
& \(3b\) & [2600, 3200] &   0 &  0.61 & 0.16  & 2.37  & 0.76 \\
& \(4b\) & [2600, 3200] &   0 & 0.042 & 0.055 & 0.72 & 0.35 \\
\cline{1-8}
\multirow{2}{*}{5000}  & \(2b\) & [4100,5200]  &   0 &  0.65 & 0.23  & 2.5  & 1.1  \\
& \(3b\) & [4200,5200]  &   0 & 0.012 & 0.034 & 1.09 & 0.29 \\
\bottomrule
\end{tabular}
\end{center}
\end{table}
 
\begin{table}[tbp]
\begin{center}
\caption{Boosted signal region data, estimated background, and signal event yields in \mhh windows containing roughly 90\% of each signal, for representative spin-2 mass hypotheses.
The signal is normalized to the overall expected limit on its cross-section; its uncertainties are evaluated by adding all individual components in quadrature.
The background yields and uncertainties are evaluated after a background-only fit to the data.
The \(4b\) category is not used for \(m(\GKK) > \SI{3}{\TeV}\) and the \(2b\) category is not used for \(m(\GKK) < \SI{2}{\TeV}\).
}
\label{tab:boosted-yields-spin2}
\begin{tabular}{lcc S[table-format=4.0] S[table-format=4.2]@{\(\;\pm\;\)}S[table-format=2.4] S[table-format=4.2]@{\(\;\pm\;\)}S[table-format=1.4]}
\toprule
\(m(\GKK)\) [\GeV]   & Category & \mhh range [GeV] & \multicolumn{1}{c}{Data} & \multicolumn{2}{c}{Background model} & \multicolumn{2}{c}{Spin-2 signal model} \\
\midrule
\multirow{2}{*}{1000} & \(3b\) &  ~~[900, 1400] & 1282 & 1298     & 29    & 12.0 & 2.1 \\
& \(4b\) &  ~~[900, 1500] &   72 &   71.0   & 6.9   & 4.14 & 0.80 \\
\midrule
\multirow{3}{*}{2000} & \(2b\) & [1600, 2400] &  223 &  221.0   & 8.4   & 3.7  & 1.2 \\
& \(3b\) & [1600, 2300] &   38 &   33.5   & 1.7   & 7.1  & 2.1 \\
& \(4b\) & [1600, 2300] &    3 &    2.15  & 0.26  & 3.2  & 1.3 \\
\midrule
\multirow{3}{*}{3000} & \(2b\) & [2300, 3400] &   19 &   16.5   & 1.2   & 2.44 & 0.93 \\
& \(3b\) & [2200, 3300] &    3 &    2.81  & 0.42  & 2.79 & 0.78 \\
& \(4b\) & [1900, 3300] &    1 &    0.63  & 0.17  & 0.82 & 0.39 \\
\midrule
\multirow{2}{*}{5000} & \(2b\) & [2400, 5300] &   18 &    13.6  & 1.1   & 3.5  & 1.4 \\
& \(3b\) & [1600, 5200] &   41 &    35.5  & 1.9   & 2.06 & 0.54 \\
\bottomrule
\end{tabular}
\end{center}
\end{table}
 
\FloatBarrier
% End of text imported from the .//sections/boosted.tex input file
% The next lines are included from the .//sections/results.tex input file
\section{Combined results and interpretations}
\label{sec:results}
 
For each signal model, the hypothesis of the presence of a signal is tested using a profile likelihood ratio~\cite{Cowan:2010js}.
The likelihood fit is carried out in bins of corrected \mhh for the resolved channel and in bins of \mhh for the boosted channel.
In the resolved channel, data from 2016, 2017, and 2018 are included separately in a simultaneous fit.
In the boosted channel, data from the \(2b\), \(3b\), and \(4b\) signal regions are included separately in a simultaneous fit.
The relative contribution of each \(b\)-tagging category to the sensitivity varies significantly with the resonance mass.
At very high masses, the track-jets associated with one Higgs boson candidate tend to be highly boosted and are often not reconstructed as individual jets.
Therefore, the \(4b\) category is used only for signal hypotheses with \(m(X/\GKK) \leq \SI{3}{\TeV}\), and the \(2b\) category is used only for signal hypotheses with \(m(X/\GKK) \geq \SI{2}{\TeV}\).
For resonance masses in the range \SI{900}{\GeV}--\SI{1.5}{\TeV}, the resolved and boosted channels are fitted simultaneously.
 
The likelihood function used to construct the test statistic has a standard form, consisting of a product of Poisson distributions for the yields in each bin
and constraint functions for nuisance parameters describing systematic uncertainties.
For uncertainties due to the limited sample size in data or MC simulation, the constraint is a Poisson distribution.
For all other systematic uncertainties, the constraint is a Gaussian distribution.
Any systematic uncertainty which is treated as uncorrelated between different regions or bins has a separate independent nuisance parameter for each of them.
Uncertainties in the luminosity and signal modeling are treated as fully correlated between the resolved and boosted channels.
All other uncertainties in the background model are treated as uncorrelated between the resolved and boosted channels.
The statistical model is implemented using HistFactory~\cite{Histfactory}.
 
The global significance is evaluated according to the procedure detailed in Ref.~\cite{Gross:2010qma}.
Pseudo-experiments are generated from the background-only model that was fitted to data, and used to construct a local \(p\)-value distribution as a function of the resonance mass.
The number of level crossings below a reference level of \(p=0.5\) is used together with the local \(p\)-value to compute a global \(p\)-value.
The most significant excess is found for a signal mass of \SI{1100}{\GeV}.
The local significance of this excess is \(2.3 \sigma\) for the spin-0 signal model and \(2.5 \sigma\) for the spin-2 signal model.
Its global significance is \(0.4 \sigma\) for the spin-0 signal model and \(0.8 \sigma\) for the spin-2 signal model.
 
\begin{figure}[tbp]
\centering
\subfloat[]{
\label{fig:limits-spin0}
\includegraphics[width=0.65\textwidth]{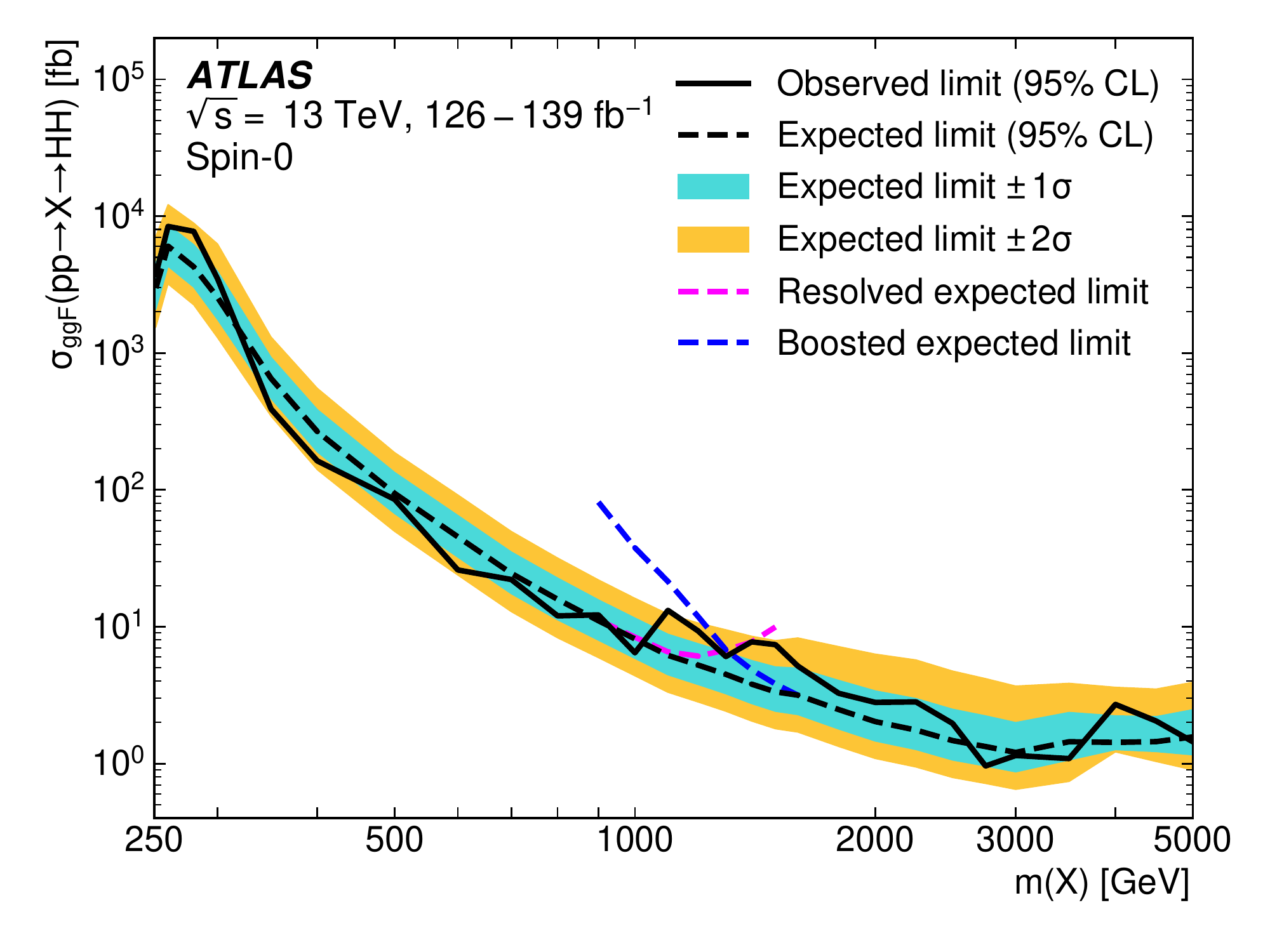}
}\\
\subfloat[]{
\label{fig:limits-spin2}
\includegraphics[width=0.65\textwidth]{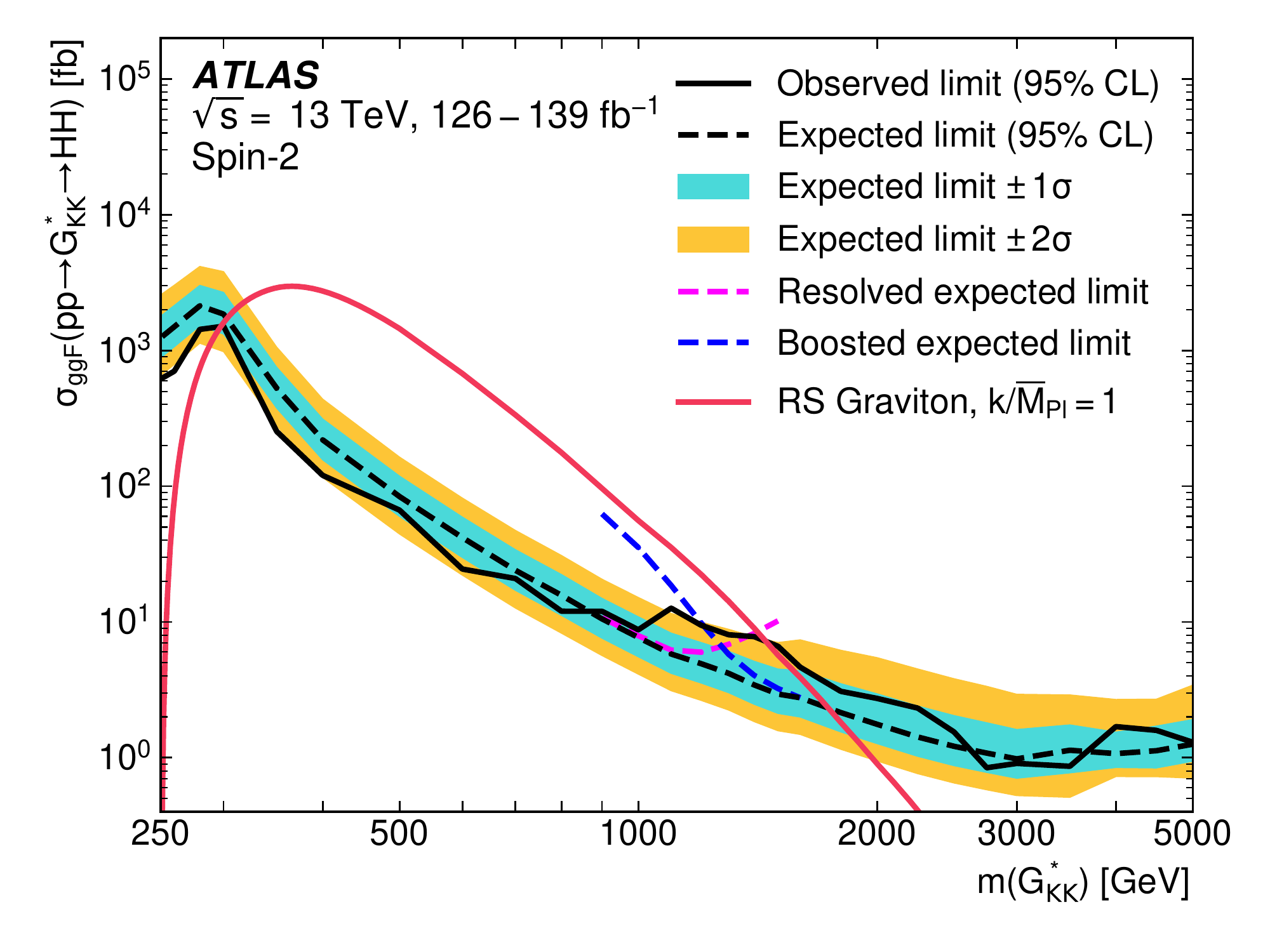}
}
\caption{\label{fig:limits}
Expected (dashed black lines) and observed (solid black lines) 95\% CL upper limits on the cross-section of resonant \(HH\) production
in the \protect \subref{fig:limits-spin0} spin-0 and \protect \subref{fig:limits-spin2} spin-2 signal models.
The \(\pm 1\sigma\) and \(\pm 2\sigma\) uncertainty ranges for the expected limits (colored bands) are shown.
Expected limits using each of the resolved and boosted channels individually (dashed colored lines) are shown.
The theoretical prediction for the bulk RS model with \(k/\overline{M}_{\mathrm{Pl}} = 1\)~\cite{Oliveira:2014kla} (solid red line) is shown;
the decrease below \SI{350}{\GeV} is due to a sharp reduction in the \(\GKK \to HH\) branching ratio.
The nominal \(H \to b\bar{b}\) branching ratio is taken as 0.582.
}
\end{figure}

Upper limits on the cross-section of resonant Higgs boson pair production via gluon--gluon fusion \(\left(\sigma_\text{ggF}\right)\) are set in each of the benchmark models.
These are based on the \CLs method~\cite{Read:2002hq}, where a cross-section value is considered excluded at the 95\% confidence level (CL) when \CLs is less than 0.05.
For signal masses up to \SI{3}{\TeV}, the limits are computed using asymptotic formulae~\cite{Cowan:2010js}.
At higher masses, the asymptotic approximation is inaccurate, so the limits are instead computed by sampling pseudo-experiments.
The results are shown in Figure~\ref{fig:limits}.
The theoretical prediction for the bulk RS model with \(k/\overline{M}_{\mathrm{Pl}}\) = 1 is also shown; this is taken from Ref.~\cite{Oliveira:2014kla}.
This model is excluded for masses between \SI{298}{\GeV} and \SI{1460}{\GeV}.
The expected mass exclusion range is from \SI{304}{\GeV} to \SI{1740}{\GeV}.
The difference between the limits on the spin-0 and spin-2 signal models at low mass is primarily due to the fact that the spin-2 model predicts a much broader corrected \mhh distribution.
In particular, the spin-2 signals with masses below \SI{300}{\GeV} are sensitive to a small deficit in the data between 350 and \SI{400}{\GeV}, while the spin-0 signals with masses below \SI{300}{\GeV} are not.
 
The impacts of the most important systematic uncertainties are shown in Table~\ref{tab:syst-summary}.
In order to compute these numbers, the limit-setting procedure is repeated, but with the nuisance parameters in question held fixed to their best-fit values instead of being allowed to vary within an uncertainty.
The resulting expected limit is an approximation of how much the sensitivity of the search would be improved if the ``true values'' of those parameters were known exactly.
Uncertainties originating from the limited sample size in any data region are not considered ``systematic'' for the purposes of this evaluation, and no corresponding fit parameters are held fixed.
For all signal mass hypotheses, statistical uncertainties are dominant.
At low masses, uncertainties in the shape of the background \mhh distribution from the data-driven estimate also contribute significantly.
These uncertainties get considerably constrained (typically by a factor of 2--3) by the fit.
Detector and theoretical uncertainties have only a very small impact on the sensitivity of the search.
 
\begin{table}[tbp]
\begin{center}
\caption{Impacts of the main systematic uncertainties on the expected 95\% CL upper limits on the signal cross-section for four illustrative values of \(m(X)\).
These are defined as the relative decrease in the expected limit when each relevant nuisance parameter is held fixed to its best-fit value instead of being assigned an uncertainty.
The spin-0 signal model is used here.
}
 
\begin{tabular}{l S[table-format=2.1] S[table-format=2.1]S[table-format=2.1]S[table-format=2.1]}
\toprule
\multirow{2}{*}{Uncertainty category} & \multicolumn{4}{c}{Relative impact [\%]}\\
& \multicolumn{1}{c}{\SI{280}{\GeV}} & \multicolumn{1}{c}{\SI{600}{\GeV}} & \multicolumn{1}{c}{\SI{1600}{\GeV}} & \multicolumn{1}{c}{\SI{4000}{\GeV}}\\
\midrule
Background \mhh shape        & 12.5 &  8.7 & 1.1 & 1.0\\
Jet momentum/mass scale      &  0.6 &  0.1 & 1.2 & 1.7\\
Jet momentum/mass resolution &  2.1 &  1.5 & 7.1 & 7.8\\
\(b\)-tagging calibration    &  0.7 &  0.4 & 2.1 & 7.0\\
Theory (signal)              &  0.6 &  0.6 & 1.4 & 1.2\\
Theory (\(t\bar{t}\) background) & \multicolumn{1}{c}{N/A} & \multicolumn{1}{c}{N/A} & 0.5 & 0.2\\
\midrule
All systematic uncertainties & 15.9 & 10.9 & 13.4 & 15.6\\
\bottomrule
\end{tabular}
\label{tab:syst-summary}
\end{center}
\end{table}
% End of text imported from the .//sections/results.tex input file
\FloatBarrier
% The next lines are included from the .//sections/conclusion.tex input file
\section{Conclusion}
 
A search for resonant pair production of Higgs bosons in the \(b\bar{b}b\bar{b}\) final state was carried out
using up to \SI{139}{\ifb} of LHC \(pp\) collision data collected by the ATLAS detector at \(\sqrt{s} = \SI{13}{\TeV}\).
Results are reported for the resolved channel, where each \bbbar pair is reconstructed as two separate small-\(R\) jets,
and the boosted channel, where each \bbbar pair is reconstructed as a single large-\(R\) jet.
The sensitivity of this analysis is improved relative to previous searches by using more sophisticated background modeling techniques, machine-learning methods, and variable-radius track-jets with optimized \(b\)-tagging in addition to the full ATLAS Run 2 dataset.
The expected upper limits on the cross-section are reduced relative to the previous ATLAS search in this final state by approximately 20\% at low resonance masses and more than 80\% at high masses.
This search also covers resonance masses in the range from 3 to \SI{5}{\TeV} for the first time.
 
No significant evidence of a signal is observed.
Upper limits are set on the cross-section of resonant Higgs boson pair production for two benchmark models:
a generic narrow spin-0 resonance, and a spin-2 graviton in the context of a bulk Randall--Sundrum model with \(k/\overline{M}_{\mathrm{Pl}}\) = 1,
both of which are assumed to be produced via gluon--gluon fusion.
The bulk Randall--Sundrum model is excluded for graviton masses between \SI{298}{\GeV} and \SI{1460}{\GeV}.
% End of text imported from the .//sections/conclusion.tex input file
 
\section*{Acknowledgments}
% The next lines are included from the .//acknowledgements/Acknowledgements.tex input file

We thank CERN for the very successful operation of the LHC, as well as the
support staff from our institutions without whom ATLAS could not be
operated efficiently.
 
We acknowledge the support of
ANPCyT, Argentina;
YerPhI, Armenia;
ARC, Australia;
BMWFW and FWF, Austria;
ANAS, Azerbaijan;
SSTC, Belarus;
CNPq and FAPESP, Brazil;
NSERC, NRC and CFI, Canada;
CERN;
ANID, Chile;
CAS, MOST and NSFC, China;
Minciencias, Colombia;
MEYS CR, Czech Republic;
DNRF and DNSRC, Denmark;
IN2P3-CNRS and CEA-DRF/IRFU, France;
SRNSFG, Georgia;
BMBF, HGF and MPG, Germany;
GSRI, Greece;
RGC and Hong Kong SAR, China;
ISF and Benoziyo Center, Israel;
INFN, Italy;
MEXT and JSPS, Japan;
CNRST, Morocco;
NWO, Netherlands;
RCN, Norway;
MEiN, Poland;
FCT, Portugal;
MNE/IFA, Romania;
JINR;
MES of Russia and NRC KI, Russian Federation;
MESTD, Serbia;
MSSR, Slovakia;
ARRS and MIZ\v{S}, Slovenia;
DSI/NRF, South Africa;
MICINN, Spain;
SRC and Wallenberg Foundation, Sweden;
SERI, SNSF and Cantons of Bern and Geneva, Switzerland;
MOST, Taiwan;
TAEK, Turkey;
STFC, United Kingdom;
DOE and NSF, United States of America.
In addition, individual groups and members have received support from
BCKDF, CANARIE, Compute Canada and CRC, Canada;
COST, ERC, ERDF, Horizon 2020 and Marie Sk{\l}odowska-Curie Actions, European Union;
Investissements d'Avenir Labex, Investissements d'Avenir Idex and ANR, France;
DFG and AvH Foundation, Germany;
Herakleitos, Thales and Aristeia programmes co-financed by EU-ESF and the Greek NSRF, Greece;
BSF-NSF and GIF, Israel;
Norwegian Financial Mechanism 2014-2021, Norway;
NCN and NAWA, Poland;
La Caixa Banking Foundation, CERCA Programme Generalitat de Catalunya and PROMETEO and GenT Programmes Generalitat Valenciana, Spain;
G\"{o}ran Gustafssons Stiftelse, Sweden;
The Royal Society and Leverhulme Trust, United Kingdom.
 
The crucial computing support from all WLCG partners is acknowledged gratefully, in particular from CERN, the ATLAS Tier-1 facilities at TRIUMF (Canada), NDGF (Denmark, Norway, Sweden), CC-IN2P3 (France), KIT/GridKA (Germany), INFN-CNAF (Italy), NL-T1 (Netherlands), PIC (Spain), ASGC (Taiwan), RAL (UK) and BNL (USA), the Tier-2 facilities worldwide and large non-WLCG resource providers. Major contributors of computing resources are listed in Ref.~\cite{ATL-SOFT-PUB-2021-003}.
 
% End of text imported from the .//acknowledgements/Acknowledgements.tex input file
 
\clearpage
\printbibliography

\clearpage
\input{atlas_authlist}

\end{document}

%% file: atlas_authlist.tex
% ATLAS Collaboration author list
% Reference date of HDBS-2018-41 is 2021-05-19
% Author list last updated on date 15-MAR-22
% Data extracted on 15-Mar-2022 for paper reference HDBS-2018-41
% at 1:39pm
 
\begin{flushleft}
\hypersetup{urlcolor=black}
{\Large The ATLAS Collaboration}

\bigskip

\AtlasOrcid[0000-0002-6665-4934]{G.~Aad}$^\textrm{\scriptsize 99}$,    
\AtlasOrcid[0000-0002-5888-2734]{B.~Abbott}$^\textrm{\scriptsize 124}$,    
\AtlasOrcid[0000-0002-7248-3203]{D.C.~Abbott}$^\textrm{\scriptsize 100}$,    
\AtlasOrcid[0000-0002-2788-3822]{A.~Abed~Abud}$^\textrm{\scriptsize 34}$,    
\AtlasOrcid[0000-0002-1002-1652]{K.~Abeling}$^\textrm{\scriptsize 51}$,    
\AtlasOrcid[0000-0002-2987-4006]{D.K.~Abhayasinghe}$^\textrm{\scriptsize 91}$,    
\AtlasOrcid[0000-0002-8496-9294]{S.H.~Abidi}$^\textrm{\scriptsize 27}$,    
\AtlasOrcid[0000-0001-5329-6640]{H.~Abramowicz}$^\textrm{\scriptsize 157}$,    
\AtlasOrcid[0000-0002-1599-2896]{H.~Abreu}$^\textrm{\scriptsize 156}$,    
\AtlasOrcid[0000-0003-0403-3697]{Y.~Abulaiti}$^\textrm{\scriptsize 5}$
% ...
\bigskip
\\

% ...
$^{5}$High Energy Physics Division, Argonne National Laboratory, Argonne IL; United States of America.\\
% ...
$^{27}$Physics Department, Brookhaven National Laboratory, Upton NY; United States of America.\\
% ...
$^{34}$CERN, Geneva; Switzerland.\\
% ...
$^{51}$II. Physikalisches Institut, Georg-August-Universit\"{a}t G\"ottingen, G\"ottingen; Germany.\\
% ...
$^{91}$Department of Physics, Royal Holloway University of London, Egham; United Kingdom.\\
% ...
$^{99}$CPPM, Aix-Marseille Universit\'e, CNRS/IN2P3, Marseille; France.\\
$^{100}$Department of Physics, University of Massachusetts, Amherst MA; United States of America.\\
% ...
$^{124}$Homer L. Dodge Department of Physics and Astronomy, University of Oklahoma, Norman OK; United States of America.\\
% ...
$^{156}$Department of Physics, Technion, Israel Institute of Technology, Haifa; Israel.\\
$^{157}$Raymond and Beverly Sackler School of Physics and Astronomy, Tel Aviv University, Tel Aviv; Israel.\\
% ...

\end{flushleft}

% Created with Glance <Atlas.Glance@cern.ch>